\documentclass[12pt,preprint]{aastex}

\slugcomment{To appear in the Astronomical Journal, March 2008} 
\shorttitle{The Early Spectra of Eta Carinae}
\shortauthors{Humphreys et al.}  

\begin{document}


\title{The Early Spectra of Eta Carinae 1892 to 1941 and the Onset of Its High Excitation
Emission Spectrum}


\author{Roberta M. Humphreys, Kris Davidson and Michael Koppelman}
\affil{Astronomy Department, University of Minnesota,
    55455\\ roberta@umn.edu}

\begin{abstract}
The observed behavior of $\eta$ Car from 1860 to 1940 has not been
considered in most recent accounts, nor has it been
explained in any quantitative model.  We have used modern
digital processing techniques to examine Harvard objective-prism
spectra made from 1892 to 1941.  Relatively high-excitation
\ion{He}{1} $\lambda$4471 and [\ion{Fe}{3}] $\lambda$4658 emission,
conspicuous today, were weak and perhaps absent throughout
those years.  Feast et al.\ noted this qualitative fact for other
pre-1920 spectra, but we quantify it and extend it to
a time only three years before Gaviola's first observations
of the high-excitation features.  Evidently the supply of
helium-ionizing photons ($\lambda < 504$ {\AA}) grew rapidly
between 1941 and 1944. {\it The apparent scarcity of such far-UV
radiation before 1944 is difficult to explain in models that employ
a hot massive secondary star,} because no feasible dense wind or obscuration by dust would
have hidden the photoionization caused by the proposed companion during
most of its orbital period. We also discuss the qualitative
near-constancy of the spectrum from 1900 to 1940, and $\eta$ Car's
photometric and spectroscopic transition between 1940 and 1953.
\end{abstract}




\section{Introduction}

Eta Carinae is famous for its many superlatives as the most massive, most luminous 
star in our region of the Milky Way and for its ``great eruption'' (1837-1858) more 
than 160 yrs ago, when it ejected 6 M$_{\odot}$ or more (Davidson \& Humphreys 1997, 
Morris et al 1999, Smith et al 2003a) forming its 
bipolar Homunculus Nebula. $\eta$ Car is also our closest and best studied example 
of a ``supernova impostor'', but there are still numerous  unsolved questions 
about the cause of the great eruption and the star's subsequent and on-going recovery.
One of the most outstanding is its continuing instability and what happened 
when $\eta$ Car experienced two post-eruption  episodes of
spectroscopic and photometric change: the second eruption (1888 - 1895) and
again from  about 1941 to 1952 when it  rapidly brightened twice and the first mention of the high
excitation emission lines appears in the astronomical literature (Gaviola 1953).  
The latter event has received little attention,
although it may be fundamental to understanding the star's process of recovery.

It is now well known that $\eta$ Car experiences a spectroscopic/photometric 5.5 year
cycle (Whitelock et al. 1994, Damineli 1996) . Modern groundbased spectra 
normally show high excitation emission lines, notably of He I, [Fe III], [Ne III], etc. 
sometimes referred to as the excitation maximum
or ``high state''. During a spectroscopic event
or excitation minimum (low state) however, the high excitation lines weaken and even
disappear. These ``events'' apparently last from a few weeks to a few months, and were 
noticed by several  authors before 1996 (Gaviola 1953, Thackeray 1967, Rodgers \& Searle 1967, Viotti 1968, Whitelock et al.
1983, Zanella et al. 1984, Bidelman et al. 1993). The 5.5 year spectroscopic cycle has been 
confirmed beginning with Gaviola's (1953) first recorded 
evidence for a spectroscopic change in 1948 up to the most recent events in 1998.0 and 2003.5
(Damineli et al. 2000, Davidson et al. 2000, 2005).

Feast, Whitelock and Marang (2001) examined the spectra of $\eta$ Car obtained at the
Radcliffe Observatory in South Africa from 1947 to 1974 and confirmed the cycle as expected.  However, 
 they also reported  that the earliest spectra from 1899 to 1919 never showed
 the high excitation state.
Gaviola (1953) had also noted that the He I emission lines were not present in the
 spectra described by Moore and Sanford (1913, 1916) and Lunt (1919). 
The plates examined by Feast et al. (2001)  included two  from 1912-1914  described by 
Moore and Sanford plus an additional plate from 1914 and two others discussed by Lunt. 
Feast et al.  also discussed Gill's early spectrum (Gill 1899) and reexamined his measurements, but 
the plate could not be located. 
Thus they examined five plates in total which they said were ``of excellent quality'', looking 
specifically for the He I line at 4471{\AA} which should have been easily discernible. 

{\it Therefore from approximately 1899 to 1919 the star's spectrum  was never in the high excitation state,
and there were  no spectroscopic events as we understand them now.} 
The 5.5 yr cycle has become an important part of $\eta$ Car's expected behavior,
and has been variously attributed to 1) the presence of a hot secondary star which is eclipsed by 
$\eta$ Car (Damineli et al. 2000)
blocking the UV radiation for the high excitation lines, 2)  a disturbance in the wind and mass ejection
(Zanella et al. 1984, Davidson 1999, Davidson et al.  2000), or 3) a combination of the two (Davidson 2005).
The absence of high excitation lines and the spectroscopic cycle in these early spectra naturally 
raises questions about the origin of the cycle and when and why  the high excitation lines first appeared.

The spectroscopic record from 1920 to the time of Gaviola's spectral series beginning
in 1944 was missing in the Feast et al. survey. However, the famous Harvard plate collection 
contains a series of objective prism plates of the 
Southern sky covering 1891 to 1951. We examined all of the available plates with spectra
of $\eta$  Car at the Harvard-Smithsonian Center for Astrophysics (CfA). In 2003 the staff at CfA  had  
just begun a trial program to digitize the Harvard plate collection, and  we were permitted to select several
representative spectra for digitization. 

In the next section, we describe the Harvard spectra of $\eta$ Car obtained from 1892 to 1941, and our rectification,
calibration and extraction process. In \S{3} and \S{4}, respectively, we discuss the earliest spectra
from the time of the second eruption, and those from 1902 to 1941.  
Measurements of the digitized spectra and comparison with recent groundbased spectra
show  that  {\it He I $\lambda$ 4471 emission was absent or only marginally present.} If present 
at all, it  was 
{\it much weaker} than
reported by Gaviola and insignificant compared with current spectra. Similarly, the [Fe III] 
lines in these spectra are either
absent or  marginal,  and there were apparently no spectroscopic events. This result has serious
implications for the origin of the He I and the high excitation lines in $\eta$ Car discussed 
in the rest of this paper. In \S{5}  
we describe the 
the apparent changes in $\eta$ Car's wind over more than 50 yrs and the onset of the high 
excitation emission during the  period of rapid  
 photometric change from $\sim$ 1941 to 1952. We review the He I emission
problem in \S{6} including  the  possible sources of present day He I emission  in $\eta$ Car and its absence prior to 1944.  
Our conclusions are summarized in the last section.

\section{ The  Harvard Objective Prism Spectra} 

The Harvard plate collection includes spectra of $\eta$ Car in
three different series of objective prism plates, listed in Table 1.
 The most extensive and most useful is the {\it X} series
obtained with a single prism on the 13-inch Boyden refractor. Spectra of $\eta$ Car were obtained 
with this telescope in 1892 -- 1926 and 1937 -- 1941. The telescope
  was moved
from Arequipa, Peru to Bloemfontein, South Africa in 1930 where it was used until 1951.
These spectra thus form a fairly uniform set  obtained with the same instrument over 
 nearly 50 yrs., although it is unfortunate that $\eta$ Car was not  observed after 1941 
when the star underwent a major photometric and spectroscopic change.   

We examined all of the {\it X} series plates that included spectra of $\eta$ Car plus a few very
early spectra in the {\it A} and {\it B} series. These plates are listed in the 
Appendix (Table A1)  in chronological order with notes on the spectra. Figure 1 shows the dates 
of the spectra superposed on a partial light curve of $\eta$ Car  together with the times 
of the early spectra from the Radcliffe Observatory (Feast et al. 2001) 
and Gaviola's spectra (Gaviola 1953). Figure 2 shows their distribution with phase
in the 5.5 yr spectroscopic cycle.  

The highest  quality spectra and the best of those representative of different times or phases in the 
spectroscopic cycle  were selected for digitization at CfA and are identified
in Table A1.  They were digitized with a Umax Power Look 3000 scanner
with 1200 bpi and a 14 bit grayscale. Figure 3 shows one example, an unusually long exposure
that clearly shows other stars around $\eta$ Car.  The plate scale was 42 arcsec/mm while
the spectral dispersion was about 28 {\AA}/mm at $\lambda$ = 4000 {\AA}
and 64 {\AA}/mm at 5000 {\AA}.   The separation between 4000 and
5000 {\AA} was about 24 mm, equivalent to an angular distance
of almost 17{\arcmin}. 
The digitized pixel values represent photographic densities, which
would be very difficult to calibrate since there were complicated
variations in exposure times, photographic response, etc.
The analysis in this paper, however,  does not depend on quantitative
relations between pixel values and radiation flux.

Many of the objective prism spectra have non-uniform guiding. The earliest spectra are  
skewed on the plate, and many of the later spectra have an obvious  curvature (Figure 3).  
Correcting for these problems and converting each spectrogram to a one-dimensional 
spectrum was a non-trivial exercise. We first rotated the image so that
$\eta$ Car's spectrum extended accurately
along image rows.  As Fig.\ 4a shows, the resulting spectrogram,
consisting of several dozen digitized image rows,  was typically
distorted by non-uniform
guiding.  Careful analysis of the emission lines showed small-scale
irregularities -- e.g., abrupt changes in tracking rate or
direction -- that are not all obvious to the eye.  Using an
iterative cross-correlation technique to find the best horizontal
offset for each image row, we straightened the spectrum as shown
in Figs.\ 4b and 4c.  We  discarded wavelengths outside the
range 3700--5300 {\AA}.  Finally, we combined the rows by the
following procedure which eliminated most scratches and localized
defects:   (1) Each digitized pixel value was divided by the
average value along its row.  (2) Then, in each column we rejected
high and low values, typically those outside the 15th and 85th
percentile.  (3) Finally we took a weighted average of the remaining
pixels in the column, where each weight was based on the average
value along that pixel's row.  In principle the  non-linear
photographic response makes some of these steps questionable;  but
in practice they worked quite well.  Most scratches and spots on
the plates -- even some alarmingly conspicuous ones -- had almost
no effect on the resulting one-dimensional spectra,  and the
exceptions were easy to identify.   Our procedure also reduced the
effects of instances where unsteady tracking caused a row to be
seriously over or underexposed, or where the telescope
accidentally lurched along the dispersion direction \footnote{FITS images of the
digitized spectra and examples of the rectified spectra plus 
the wavelength calibrated tracings are available at http://etacar.umn.edu/download/EarlySp and with the electronic edition of the Astronomical Journal.}.

An accurate wavelength calibration was necessary for judging the
presence or absence of \ion{He}{1} $\lambda$4471, discussed later. 
 The only feasible calibration data were emission
lines in the spectrum of $\eta$ Car itself.  Practically all of these
were unresolved blends, usually of  \ion{Fe}{2} and [\ion{Fe}{2}]
lines which originate at various distances from
the star;  and in principle the line ratios within each blend
depend on the spatial coverage and evolve with time as the
ejecta densities progressively decrease.  Therefore, as reference
standards we used wavelengths measured by Gaviola from spectra obtained in the 1940's. 
His were the earliest  and most complete  
published data with sufficient accuracy and were obtained  relatively close to the  time 
to when many of the
objective prism plates were observed.
His data were photographic, and the spectra  were ``slitless'', 
with a comparison spectrum exposed before and after through a superimposed slit.
Like the objective prism spectra, they include the entire star plus nebula. 
In each of the selected spectrograms we measured 12 to 22 emission
features (blends) in the wavelength range 4173--5018 {\AA}, not
using any hydrogen lines or blends that might include helium
emission.   Then we fit the measurements to a ``Hartmann formula,''
\begin{displaymath}
   x  =  a + \frac{b}{(\lambda - c)} ,
\end{displaymath}
where we determined the adjustable parameters $a$, $b$, and $c$
separately for each individual case.  This expression
resembles  the prism dispersion much better than a three-parameter
polynomial does, and we found that a fourth parameter did not
materially improve the fit.  In a simple model the wavelength
asymptote  $c$ should have been a constant for all the plates;
in fact the best-fit values were all in the range 1930--2070 {\AA},
acceptably similar, under the circumstances.

The resulting wavelength calibrations were remarkably good.  The
r.m.s.\ deviation of individual data points from the three-parameter
fit was typically $\pm$0.25 {\AA}, and about $\pm$0.15 {\AA}
for the best spectrogram, X16614.  The {\it formal\/}
one-sigma uncertainty in the wavelength fit around 4500 {\AA} was
therefore only about $\pm$0.1 {\AA} (7 km s$^{-1}$) for most cases
and $\pm$0.04 {\AA} for X16614.  
To some extent {\it systematic\/} effects are constrained by a test noted below.  
On the other hand, since 12--22 points were used to
adjust a three-parameter formula with an appropriate
functional form,  the calibration errors for wavelength
{\it differences\/} between features in any 100-{\AA} interval near
the middle of the range are presumably much smaller than
0.1 {\AA}.

To forestall ambiguities in wavelength values
quoted later in this paper, we shifted each spectrum to the
laboratory wavelength system ``in air'' (STP).   We assumed a Doppler velocity
of $-30$ km s$^{-1}$, based on the measurements for the metallic  emission lines
by the early observers (Gaviola 1953, Lunt 1919).
({\it Caveat:\/}  In $\eta$ Car the hydrogen lines can
have significantly different Doppler shifts than \ion{Fe}{2}, [\ion{Fe}{2}],
\ion{Ni}{2}, etc.)

The overall wavelength measurement errors are crucial to the discussion in
Section 4 below.  In order to assess the likely
uncertainty, we measured the wavelength of
[\ion{Fe}{2}] $\lambda$4287.40 in 12 spectrograms,
using the same procedures as we did for the lines
discussed in Section 4.   This is the only strong
feature that is dominated by a single, narrow,
well-identified spectral line in these data, so
(unlike the blends) there is no doubt about its correct
laboratory wavelength.  Since we did not give it extra
weight in the wavelength calibration, it is a valid
reference for testing our measurement technique.
We measured the near-peak part of the line, essentially
the centroid wavelength in the upper 20\% of the feature's
net height;  this quantity is well-suited to the non-linear
photographic response.  Among the 12 spectra, the r.m.s.\
deviation from 4287.40 {\AA} was $\pm$0.22 {\AA}.
Several of them, however, were recognizably of low
quality, with asymmetric or blurred emission features.
When we omitted three obviously poor spectra, the
r.m.s.\ wavelength error improved to 
$\pm$0.13 {\AA}.

This estimate is smaller than the r.m.s.\ data-point deviation
quoted above for the wavelength calibration -- which is not very
surprising, since wavelengths {\it assumed\/} for that procedure
were inexact, representing blends rather than individual spectral
lines.  (Errors in the assumed wavelengths of blends presumably
averaged out in the calibration process, since many independent
blends were used as data points.)   Thus, in the absence
of additional information, $\pm$0.13 {\AA} is the best error
estimate we can make for a careful measurement in a satisfactory
spectrogram.  We emphasize one major exception, however:
{\it spectrum X16614 was conspicuously better.\/}  It had much
sharper features than most of the others (See Fig. 11), and produced formally
better wavelength-calibration statistics.
Its exposure level, guiding, etc.  must have been fortuitously
optimal.  Judging from all the considerations outlined above,
the standard error for a wavelength measurement in X16614 is
 $\pm$0.1 {\AA} or better;  informally we suspect that
$\pm$0.08 {\AA} would be a realistic estimate for well-defined
spectral features in X16614. We measured 4287.34{\AA} for the [Fe II] 
$\lambda$4287.40 line in X16614.

\section{The Earliest Spectra 1892 -- 1898}

The first spectrum of $\eta$ Car was a visual
observation by Le Sueur (1870) who described  strong emission lines,
and the first photographic spectrogram was obtained on May 15, 1892
with the Harvard 13-inch Boyden telescope at Arequipa.
The spectrogram from June 2 1893, X4709, is of especially good quality and has
has been described by several  authors including Cannon (1901), Bok (1930),
Hoffleit (1933), Whitney (1952)  and most recently by Walborn and Liller (1977).
Unlike spectra observed after 1895,  it shows a well developed absorption line spectrum  
resembling an F-type supergiant plus strong emission lines of hydrogen, Fe II  and [Fe II]. 
These earliest Harvard spectra from 1892 to 1895 were obtained when $\eta$ Car was in its
second or lesser eruption (1888 -- 1895). The F-type supergiant spectrum, combined with the star's
apparent brightening of about two magnitudes which lasted $\sim$ 7 yrs , suggests that this was
a classic LBV or S Doradus-type eruption with 
an  optically thick cool wind (Humphreys, Davidson \& Smith 1999, Humphreys \& Davidson 1994).

Our digitized scan of the famous 1893 spectrogram  is shown in Figure 5  with  
the rectified scan, and the extracted tracing.
A good reproduction of 
the original photographic spectrum can be seen in Walborn and Liller (1977). 
The procedure described above (\S{2}) for the wavelength calibration was 
unsuitable for  X4709 because its spectrum was  fundamentally different.  Instead,  we
used  several emission lines measured by Whitney (1952),
who used a difficult  calibration procedure based on the hydrogen lines in the
 spectra of other stars on the plate.
Thus our wavelength calibration for X4709 is not as good
as the cases described above.
Whitney  gave a comprehensive list of numerous identified absorption lines with estimated 
line intensities that are consistent with its early-to-mid-F-type supergiant classification.  
His  measured velocity of $-180$ km s$^{-1}$ for the absorption lines 
relative to the H emission lines is indicative of an expanding 
envelope  consistent with the comparatively slow winds in LBV eruptions (Humphreys \& Davidson 1994). 
Weak absorption features, attributed by  Whitney to hydrogen absorption,  
can be seen to the blue of the  H$\gamma$ and H$\delta$ emission lines and may be due to P Cygni absorption. Using our
wavelength scale, we measured velocities at the base of the absorption, relative to the 
hydrogen emission peaks,  
of $-300$ km s$^{-1}$, significantly slower than $\eta$ Car's current polar wind (Davidson \& Humphreys
1997, Smith et al. 2003b), but comparable to the polar expansion velocity in the ``little 
homunculus'' (Smith 2005) from material probably ejected in the second eruption.

The additional spectrograms obtained between 1892 and 1900 are described in Table A1, 
although most of the
spectra from this period are too weak to confidently identify the lines.
We note that the F-type absorption lines had vanished by 1895 when the second 
eruption had ended.  
Spectra at lower dispersion from 1897 and 1898 show only hydrogen emission.
Beginning with the observations in 1902, the spectra begin to show the prominent 
emission line spectrum of hydrogen, [Fe II], and Fe II that is familiar in all 
subsequent groundbased spectra of $\eta$  Car.

\section{The Spectra from 1902 to 1941} 

The Harvard spectra have good coverage during several intervals:
1902--1904, 1913--1916,  1922--1926, and 1937--1941.  The best  
spectrogram, X16614 obtained on 29 March 1938, is shown in Figure 6. 
Figures 7 and 8 show expanded regions of the same spectrum with some of the stronger lines 
identified.

For reasons given in the Introduction and discussed in \S{5} and \S{6}, it is important to
determine whether helium emission features were present.  In the
wavelength range covered by these spectra, the most suitable and strongest line is
\ion{He}{1} $\lambda$4471.5 which was also discussed by Feast et al (2001).  
Figure 9 shows an obvious emission feature or
blend centered at 4473.7{\AA} in X16614, with an expected measurement
uncertainty of better than  $\pm 0.1${\AA}
\footnote{Using a  simple interpolation between [Fe II] 4458{\AA} and the 
Fe II blend at 4490{\AA}, Humphreys and Koppelman (2005) originally estimated a line 
center of 4474.0{\AA}.}. This wavelength refers to both the peak and the centroid of the
brightest part of the feature. Four emission lines are expected between 4470{\AA} and 4475{\AA} 
(Aller \& Dunham 1966). Apart from helium $\lambda$4471.5, the feature is dominated by 
[Fe II] $\lambda$4474.9 plus
two weaker lines, Fe II $\lambda$4472.9 and [Fe II] $\lambda$4470.3.   
Gaviola (1953) states that He I $\lambda$4471.5 was about  the same 
strength as  [Fe II] $\lambda$4474.9 when his spectra were obtained in 1944-1951. If this feature were a simple blend of these two lines,
its expected central wavelength would be 4473.2{\AA}, significantly to the blue of the feature in X16614.
Based on Gaviola's eye estimates, the Fe II $\lambda$4472.9 line  was  much weaker than  
the $\lambda$4474.9 line, and  the $\lambda$4470.3 line was  blended with He I in
his spectra.
Thus the 4473.7{\AA} emission feature in X16614 is obviously a blend, and
He I $\lambda$4471 was apparently weaker than in Gaviola's spectra after 1944. 
Whether or not  it was present in X16614 depends on the relative strengths of the other 
three lines.  

For reasons noted later, our quantitative assessment of the pre-1941
high-excitation features will rely on comparisons with published
spectra obtained in 1944--1961, and not on the spectrum seen today.
Nevertheless, in Figures 9 and 10, we show X16614 together with recent groundbased data, 
because they clearly show the pertinent features and they are readily
available while tracings of older spectrograms are not.
These modern data were obtained using the GMOS-S on the Gemini
South telescope on 2007 June 11,  with exposure time 5 s, slit width
0{\farcs}5, and seeing about 1{\farcs}5.   The tracings in Figs.\
9 and 10 represent an extraction width of 1{\farcs}5 along the slit.
Because of the unusually poor seeing, in effect the extracted spectrum
samples a region almost 2{\arcsec} across, including about 60\% of the
total brightness of the star plus Homunculus Nebula (see Figs.2  and
3 in Martin et al. 2006b).  It therefore shows the broad-line
spectrum of the opaque stellar wind combined with narrower lines from
slow-moving material ejected many years ago.\footnote{
     HST/STIS data contain more detail and they do not mix the
     spectra of $\eta$ Car's wind and nearby slow ejecta together
     as ground-based observations do.  See archives at
     http://etacar.umn.edu/ and http://archive.stsci.edu/prepds/etacar. }
The vertical scale is linear for the Gemini tracing and
quasi-logarithmic for X16614, but here we are chiefly interested
in wavelengths rather than flux values.  The wavelength scale is
in the laboratory rest system for both spectra. Note that the wavelength 
correspondence between the 1938 and 2007 spectra is quite good.   

The He I and [Fe II] 4474.9{\AA} lines are easily separated in the recent Gemini spectrum. 
The peak or center of the objective prism 4473.7{\AA} emission feature is
in the trough between these two lines,  but shifted appreciably to the red with respect 
to the weighted  centroid (at 4472.9{\AA}) of the two present day emission peaks. 
If He I were present in the objective 
prism spectrum in comparable strength to the current spectrum, it would appear as a bump 
or shoulder on the blue side of the 4473{\AA} feature. If it were present in 
approximately  equal  strength to [Fe II] 4474.9{\AA}, as Gaviola reported, then the peak or
center of the blend would be at about 4473.2{\AA}, as we noted above.   
The Fe II 4472.9{\AA} line can just barely be discerned in the Gemini spectrum
as a  small bump on the red side of the He I profile which we measured at 4473.1{\AA}. Although
it is one of the weaker lines in Fe II multiplet 37,  we believe this identification is correct, because 
all of the stronger lines and one weaker line in multiplet 37 are all present in
the recent spectrum. We also identified all of the same lines in the 1938 spectrum. This Fe II line was 
thus present  then and contributed to the observed 4473.7{\AA} profile in X16614.           

Today a substantial fraction of $\eta$ Car's observed light
comes directly from the emitting regions, and the star can be
separated from its slow-moving inner eruptive ejecta at high
spatial resolution;   see Martin et al.\ (2006b) and refs.\ therein.
Several decades ago, however, the situation was qualitatively
different.  Circumstellar extinction was then so large that nearly
all of the observed visual-wavelength light escaped {\it only after
being scattered by slow-moving dust grains\/} (see Davidson and Ruiz 1975
for a typical model).  Then as now, the inner radius of the dusty
region should have been 150 to 600 AU, depending on grain details;
but optical depths were then larger than today's, and more slow-moving ejecta
were located fairly close to the star.
Thus, before about 1970 or 1980, most spectrograms of $\eta$ Car
represented a mixture of broad emission features from the stellar
wind plus narrow lines from inner slow-moving ejecta, altogether
diffusing out through a scattering nebula.  Only about 10\% of
the light escaped.  The scattering region had an apparent diameter
of the order of 2{\arcsec}, and contained a central dust-free zone
because grains could not exist close to the central star.   In view
of these circumstances, we shall use pre-1970 spectrograms, not
modern data, as reference comparisons for the old objective prism
spectra.  The most useful examples are Gaviola (1953), because most
of his spectra were obtained before 1950;  and Aller and Dunham (1966),
because they carefully calibrated the relative line intensities
visible in 1961.

Our analysis depends on wavelengths. In order to avoid possible
misunderstandings, let us make some preliminary comments about
Doppler shifts.  First, there is no reason to think that Doppler
shifts caused by scattering would especially perturb the high-excitation
\ion{He}{1} and \ion{Fe}{3} lines;
no such effect can be seen in the measurements reported by Gaviola and
by Aller and Dunham, and in general one expects the entire spectrum
produced near the star, both low- and high-excitation lines, to be
similarly affected by the scattering process.   In a related connection,
we note that Doppler shifts of more than 100 km s$^{-1}$ occur when the
central spectrum is reflected by fast outward-moving material in the
Homunculus lobes (Davidson et al.\ 2001 and older refs.\ therein).  This
fact has no effect on our measurements, because the Homunculus lobes
were relatively faint;  the objective prism plates, Gaviola's
spectrograms, and that of Aller and Dunham were all strongly
dominated by the bright central core mentioned above.  This
can be seen in those authors' measurements and in the tracings
shown by Aller and Dunham, which resemble the modern Gemini
spectra to a surprising degree.  Their Fig.\ 2 is especially
relevant because it shows the \ion{He}{1} $\lambda$4471.5 line.
Next, the present-day {\it broad\/} \ion{He}{1} $\lambda$4471.5
profile fluctuates throughout $\eta$ Car's 5.5-year spectroscopic
cycle;  its centroid wavelength is generally a little smaller than
4471.5 {\AA}, which should make the feature more, not less, evident
in the type of analysis we employ below.  Since we find a
non-detection, we do not attempt to correct for this complex
effect.   Finally, we note that most pre-1970 spectrograms
of $\eta$ Car did not perceptibly depend on their chosen slit
sizes or even the absence of a slit.   They were dominated by
the bright central core, and atmospheric seeing caused nearly
the entire core to be sampled within any normally-used slit
width.\footnote{
   The only exceptions to this statement were slit spectra of
   the outer Homunculus that avoided the central object;  but
   they are irrelevant in this paper.}

To determine whether He I emission was present in the 1938 spectrum we must  
estimate the relative strengths  of the  [Fe II], Fe II and He I lines  required to 
produce the observed blended profile.  
A detailed model of the profile would require quantitative information about the photographic response
 which is
no longer available. Instead, we have adopted  the spectrophotometric  measurements of the line 
strengths from Aller \& Dunham's (1966) high 
resolution Coud$\acute{e}$ spectra from 1961. 
Using their values we find an average wavelength of 4473.7 {\AA} for the
[\ion{Fe}{2}], \ion{Fe}{2} blend, i.e., not including
any \ion{He}{1} $\lambda$4471.5 emission.
{\it Thus  no contribution from He I $\lambda$ 4471.5 is required to produce the 
observed line center at 4473.7}.  X16614 was obtained at phase 0.22 in the
present-day 5.5 yr cycle, so it does not correspond to a spectroscopic event.   
Aller \& Dunham's spectra were obtained more that 20 yrs after X16614, so as a check, we also
used Gaviola's eye estimates of the line strengths and obtained the same result. 

The expected one sigma error for this measurement in X16614 is less than 0.1{\AA} (\S{2}). Thus allowing
for an overall uncertainty of $\pm$2$\sigma$,  the line center is between 4473.5 
and 4473.9{\AA}.
Any shift to the red would confirm no contribution  from He I. If we adopt a
line center at 4473.5{\AA} as the shortest feasible wavelength,  the He I contribution is
still  small, less than 10\% of the total blend and about 18\% as bright as the  
$\lambda$4474.9 line observed in 1961. In that case,  He I in 1938 would have been less
than   one-fourth its strength ssen in Gaviola's 1944-51 spectra when He I  and [Fe II] $\lambda$4474.9 were about equal. Based on other line ratios in 
both the 2007 Gemini data and Aller \& Dunham's
1961 spectrum, it is fair to conclude that I(He I $\lambda$4471) $<$  0.003I(H$\beta$) in 1938. 
{\it Thus we conclude that helium emission was quite weak and may have been 
virtually absent in 1938. } 

None of the other  strong high excitation lines visible in modern 
spectra are apparent in the 
1938 spectrum \footnote{A weak emission feature is present at $\lambda$4713.9, but given its
measured wavelength and the absence of any other He I lines in the spectrum, we concluded that this is not
He I $\lambda$ 4713.1, but more likely a blend of Fe II and Cr II. If it were He I $\lambda$4713, with normal helium line ratios, then we should have easily detected $\lambda$4471.}.  
The best example is   [Fe III] emission at $\lambda$4658.1 which is quite 
strong in the spectra of eta Car described since 1951 (Thackeray 1953, Aller \& Dunham 1966, Zethson 
et al 1999).  The [Fe III] emission lines are present in Gaviola's data, but notably  weaker than in later spectra.
If [Fe III] $\lambda$4658   is present in X16614, it is part of a broad complex emission profile 
from 4652{\AA} to 4674{\AA} due to several different lines (Figure 10). The red peak 
in this profile at $\lambda$4666.7 is 
due primarily 
to   Fe II $\lambda$4666.75, and the profile also includes  Fe [II] $\lambda$4663.7, [Fe II] $\lambda$4664.5 and Fe II $\lambda$4670.2. 
 We measure a center for the much smaller secondary peak on the blue side at 4657.5{\AA}. In addition to a possible
contribution from the [Fe III] line, Fe II $\lambda$4657.0 and [Co II] $\lambda$4657.4 (Zethson 2001) plus lines
of Cr II are also present. 
Given the complexity of this profile, and the lack of a recognizable peak due to the [Fe III] line, 
visible in the later
spectra, it is difficult to ascertain if it is present, although  a weak feature just above the continuum 
at 4701.1{\AA} may be [Fe III] $\lambda$4701.5. Assuming that some [Fe III] emission is 
contributing to these features, then like He I, they are  much weaker, with respect to 
other emission lines, than in the later spectra.  
In the course of comparing the objective prism spectra with Gaviola's spectra and other published
line lists and intensity estimates in the literature, we noticed that beginning in 1951-52 the apparent
strength of
the high excitation emission lines appeared to increase  compared to the Fe II and [Fe II] spectrum
To document this we compared  the line ratios of He I 4471.5{\AA} relative
to [Fe II] 4474.9{\AA} and [Fe III] 4658{\AA} to  [Fe II] 4639.7{\AA} from several sources  using their
own estimates of the line strength or intensities.  These are summarized in Table 2.
The He I emission has remained relatively constant since $\sim$ 1951, while the strength
of [Fe III] has gradually increased. Decreasing gas densities  may partially explain this trend.

We adopted  the 1938 spectrum as a template or reference for comparison with all of
the other digitized spectra from 1902 to 1941. We looked for any changes in the profile 
and central wavelength of the 4473.7{\AA} emission feature as well as any other obvious
changes in the spectrum. This emission feature is present in all of the  
spectra independent of phase in the 5.5 yr cycle. In the earliest spectra (1902 - 1916),
this line was noticeably much weaker than in the 1938 spectrum and barely visible 
above the noise in some cases (Figure 11). 
This is obviously the line noted by Feast et al (2001) and seen by the early
observers in spectra from 1913 and 1919. We therefore concur with Feast et al that He I was 
not present in any significance in these spectra and that the observed feature is  a  blend of 
the Fe II and [Fe II] lines. The [Fe III] emission lines  are not present in any of these earliest
spectra. In the spectra from 1922 to 1941, the 4473.7{\AA} emission feature is virtually identical 
to the same feature in the 1938 reference spectrum with respect to the shape of the profile and the  
measured central  wavelength of the blend.  We also note that  the  [Fe III] lines 
 at 4658{\AA} and 4701{\AA} are either  absent in most of these spectra or, may be marginally present in a few, as in X16614. The related [\ion{Ne}{3}] emission lines
(Zanella et al.\ 1984) are outside the wavelength range
where these plates were sensitive. 
 
Did spectroscopic events occur before 1948?  Only a
few of the Harvard spectra were obtained near the time of an expected ``event'' (Fig. 2).
Those from May 1937, phase
0.06, show  no apparent change in the spectrum or in the 4473.7{\AA} emission feature.
Evidently, between the time of 
the last spectrum in the Harvard series from May 1941 and Gaviola's spectral series beginning 
 three years later in 1944, the high excitation lines appeared and the He I emission 
increased significantly in relative intensity.
The star also  brightened by almost a magnitude sometime between 1939  
and June 1941 (O'Connell (1956), de Vaucouleurs \& Eggen (1952)). $\eta$ Car's wind was 
apparently in a critical transition during that time. For this reason we carefully inspected
the spectra between 1938 and 1941  for any spectroscopic evidence 
and found no obvious change.

Viewing all the Harvard plates together, one especially notices how  similar all of
these spectra appear  after 1900.  They all show the same qualitatively similar
features.  Moreover, apart from He I and the strong high-excitation lines,
they closely resemble subsequent low-resolution groundbased spectra since 1950.
This is remarkable because
the Homunculus nebula tripled in size during the twentieth
century, and the object experienced more than one episode of rapid brightening.

In summary we  conclude that  {\it the helium emission lines
were quite weak, practically absent, during the period 1902--1941.\/} This applies to both the
broad and narrow components. Similarly, the high
excitation forbidden lines were either absent or very weak.  
$\eta$ Car's spectrum was thus  always in the ``low excitation state'' from 1900 to 1941. 
This fact has serious implications discussed in \S{5} and \S{6} below.

\section{Variations in the Wind of Eta Car 1870 to 1951}

Beginning with the first visual spectroscopic observation by Le Sueur (1870), 
 we can recognize four different stages defined by the appearance of $\eta$ Car's  spectrum
and its light curve; 
(1) post-eruption 1858 - 1887, (2) the second eruption 1888 - 1895, (3) quiescence $\sim$ 1900 - 1941,
and (4) the spectroscopic/photometric transition 1941 - 1952.

\subsection{Post-Eruption} 

We know very little about the physical state of the star after the great eruption except that it
faded rapidly after 1857. This is usually attributed to the formation of dust which 
probably occurred quickly after the cessation of the eruption, although the exact time when the 
eruption ended is not clear from the light curve. Le Sueur's (1870) visual observation 
about 10 years later is especially intriguing. He described five emission lines, the Fraunhofer
lines C (H$\alpha$), D, ``b group'' and F (H$\beta$) plus the ``principal green nitrogen line''.  
The Fraunhofer b group,
due to Mg I absorption in the Sun (5167 -- 5184{\AA}),  is very likely [Fe II]
5158{\AA} and Fe II 5169{\AA} emission in $\eta$ Car. The ``D'' line is especially interesting, because
although Na I D is present today in $\eta$ Car as a complex mix of absorption and emission, it is not a strong
emission line. Strong Na D emission occurs in much cooler stars. Le Sueur says the yellow or orange line is 
``very near D''. Thus it is possible  
that it was actually He I 5876{\AA} and if so it had to be at least as strong then as it is now. In current
groundbased spectra it is comparable in strength to the [N II] 5755{\AA} line which apparently wasn't
noticed by Le Sueur. It is clear from the text that this was a 
difficult observation, near the limit of what could be seen, so it is possible that He I 5876{\AA}  and  
other emission
lines were much stronger then. As this was only about 10 years after the great eruption during which $\eta$ Car
had thrown off its entire outer envelope, we can only speculate about the origin of a He I line, but it
 could either originate in a  hot wind or even from what was then its photosphere.  Walborn and Liller (1977)
earlier made these same identifications based on Le Sueur's description and  suggested that
the green nitrogen line was [Fe II] $\lambda$5018 because of Le Sueur's statement that is was not just an
extension of the nebular emission line, presumably [O III] $\lambda$5007, across the spectrum of 
$\eta$ Car \footnote{On the other hand, Schuster (1872) explicitly stated
that  ``the green line'' was near 5164 {\AA} in an arc spectrum
of nitrogen;  perhaps he meant \ion{N}{2} multiplets 35
and 37.   This would coincide with Fraunhofer b,
but Le Sueur did not explicitly say that he could distinguish
``the principle green nitrogen line'' from Fraunhofer b.
Therefore, although Walborn and Liller's suggestion still seems
most likely, it is not certain.} .          

\subsection{The Second Eruption}

The second or lesser eruption is now best understood as similar to a classical LBV or S Doradus-type eruption or
 optically
thick wind stage (Humphreys et al 1999) during which the expanded wind or ``pseudo-photosphere'' 
is relatively slow, dense,  and cool with an enhanced mass loss rate. The second eruption was not as minor as 
it may appear on the historical light curve (Figure 1). When 
corrected for a realistic estimate of the circumstellar extinction at that time,  the star's apparent
brightness would have been  about second magnitude (Figure 3 in Humphreys et al. 1999), and its expanded,
 cool envelope with its F supergiant spectrum would have reached {\it r} $\sim$ 6 AU. The star also ejected 
about 0.2 M$_{\odot}$ of slow
moving material in both the equatorial and the polar directions (Davidson et al. 2001, Ishibashi et al 2003, Smith 2005).   
Thus in both the great and second eruptions,  $\eta$ Car ejected  mass with a bipolar structure. 
The cause of this second outburst is not
known, but the  most likely explanations for LBV-type eruptions  
include the opacity modified Eddington limit, subphotospheric gravity-mode instabilities, and a super-Eddington
continuum -driven wind (See reviews by Humphreys \& Davidson 1994, Glatzel 2005, Owocki 2005, Shaviv 2005).    

\subsection{Quiescence $\sim$ 1900 to 1941}

The period from about 1900 to 1941 was one of relative quiesence for $\eta$ Car. Its apparent brightness was
basically constant at $\sim$ 8th magnitude, and as we have  described, the spectra from this period are remarkably 
similar. Weak  He I emission may have contributed  to the blended emission feature at 4473.7{\AA}
but, if so, it   was only marginally present and much weaker than observed after 1944. 
With very weak or absent He I and other  high excitation lines, the spectrum was always
in what is now called the low excitation state and there were no spectroscopic events.    

Today we make a clear distinction between emission lines produced
in $\eta$ Car's wind ($r < 15$ mas) and the features that originate
in ejecta from past outbursts ($r > 0\farcs1$); see below.
Groundbased spectra include both types together.  Before 1950,
however, this distinction would have been less clear even
if high spatial resolution had been available.  The innermost
dense ejecta were then closer to the star, and the wind may have
been denser. The observed emission therefore must have escaped
from the inner region via multiple scattering by dust grains
(Davidson \& Ruiz 1975).   All emission sources within
1{\arcsec} of the star were thus mixed together in the
emergent spectrum, as we noted in {\S4}.

The broad He I emission lines originate in $\eta$ Car's wind as discussed in \S{6} below,
while the narrow He I and high excitation forbidden lines are now known to come from
slow moving diffuse gas near the star, including the Weigelt knots within 0$\farcs$3 
(Davidson et al 1995). The relevant high-excitation gas is
presumably photoionized (Zanella et al.\ 1984;  Davidson 1999). 
In the colliding-wind/X-ray eclipse model (Pittard \& Corcoran 2002)  for
the spectroscopic events, the responsible UV radiation is assumed to come from the secondary star which is
then eclipsed by the primary during an ``event''.  Despite a claim to the contrary (Iping et al. 2005), 
there is no unambiguous 
spectroscopic evidence for the companion (Hillier et al. 2001, 2006, Martin et al. 2006a). Various observational
constraints, such as the 3000 km s$^{-1}$ wind required for the X-ray spectrum and the 
star's lack
of visibility in the spectrum, suggest that the secondary must be a hot, main sequence star, 
much less luminous than $\eta$ Car, probably 
an O-type star with initial mass of $\approx$ 40 -- 60M$_{\odot}$ (Ishibashi 2001, 
Davidson 2005). It is thus a  massive  star,
with a normal hot star wind, but far less massive than the primary. 

But if this same star is the source of the UV radiation for the  high excitation lines,  there must
be some explanation for why they were not observed for 40 years or more. 
Two possible  explanations  are  obscuration of the secondary by either dust or by an 
optically thick wind from the primary, preventing the UV radiation from reaching the inner ejecta. 
There are feasible  ways to prevent the 
{\it narrow\/} high excitation lines from originating 200--1000 AU from the
central star;  but the broad \ion{He}{1} $\lambda$4471.5 line formed in the wind is much harder
to suppress. The UV flux from one or both stars will destroy dust within 100 AU and the expected 
dust formation distance is 150--600 AU.  
The semi-major axis of the secondary's highly eccentric orbit is 17 AU 
so it is always well within the dust envelope. Therefore dust could not prevent its UV flux from
ionizing  helium in the primary's wind, and we should have detected  
moderately broad \ion{He}{1} $\lambda$4471.5 emission 
(see Fig.\ 2 in Aller and Dunham 1966).\footnote{
     As D.J. Hillier (priv.\ comm.) remarks, conceivably one might
     devise an explanation based on inhomogeneities in the dust
     distribution.  Observations since 1980 suggest that the star is
     behind a local extinction maximum with a size scale of only
     0.2{\arcsec} (Davidson et al.\ 1995;  Hillier \& Allen 1992;
     Davidson \& Humphreys 1986).  If for this reason the
     pre-1941 spectrum mainly represented emission far from the
     star, then perhaps \ion{He}{1} emission formed in the primary
     wind would have been hidden.  An explanation of this type has
     several difficulties, though.  The H$\beta$ line in the
     objective prism spectra is perceptibly broad, indicating that
     it was formed in the wind; how can line-of-sight dust hide
     \ion{He}{1} $\lambda$4471.5 but not H$\beta$?  Moreover, one
     would also need to explain why the high-excitation lines
     {\it did\/} appear in Gaviola's spectrograms in the mid-1940's.}
The Weigelt knots, for example, are today $\approx$ 400 - 500 AU from the star, but in 1900 - 1940
this slow moving material, probably ejected in the 1890's eruption (Smith \& Gehrz 1998), 
was also mostly  inside the dust formation zone. 
Thus  obscuration by dust cannot easily explain the weak or absent He I 
and  high excitation lines prior to 1944. 
An optically thick wind that would enshroud the secondary throughout 
its orbit would also not suppress the broad He I lines, see \S{6}.

\subsection{The Spectroscopic/Photometric Transition 1941 - 1952}  

Between the end of 1939 and mid 1941 $\eta$ Car  brightened about 0.8 mag on the photographic scale 
(O'Connell 1956). Only three years later, the star's spectrum showed prominent He I emission 
and other high excitation emission lines. 
It continued to brighten for several years with small oscillations reaching 6.9 to 7.1 photographic by 1952 
when de Vaucouleurs \& Eggen (1952) reported a rapid increase from  6.9 to 6.5 mag in apparent 
visual brightness, measured  
photoelectrically, in only a few weeks from February to March 1952. It is important to realize that
these measurements and eye estimates refer to the integrated light of the entire Homunculus not just
to the central object. Neither of these brightenings correspond with a spectroscopic event; the preceding events occurred in 1937.0 and 1948.1\footnote{There are two gaps in the photometric record during  the 20th century, 1915 - 1935 and 1953 - 1970. We suspect that a brightening of one
magnitude or more would have been noticed especially in the 1950-60's when Thackeray was 
observing $\eta$ Car frequently}. The object's  rapid brightening during this time was most 
likely due to dust
destruction which might  also be caused by  the increase in UV radiation required for the onset
of the He I and high excitation emission lines. As  mentioned in \S{4}, spectra obtained after 
1951-52, show an increase in the
strength of the  high excitation emission lines such as [Fe III] compared to Gaviola's spectra. 
After 1952, the integrated light of the
Homunculus  continued to brighten slowly at a fairly regular rate,  attributed to the
 expansion  of the dust envelope,  until HST/STIS spectra in 1998-99 revealed another rapid
brightening, this time of the central star, of $\sim$ 1 magnitude in only one year (Davidson et al 1999). Since then the
integrated light  
has continued to brighten more rapidly than the previous long-term trend (Martin \& Koppelman 2004, Martin 
et al. 2006b)

The colliding-wind binary model provides a reasonable  working hypothesis for the X-ray light curve and the 
5.5 yr period. However,  main sequence hot stars with their  
radiatively driven fast winds are not expected to experience sudden increases in UV radiation
implying a drastic change in the wind and mass loss rate or an increase in the star's 
surface temperature. 
Smith et al (2003b) proposed an explanation for the increase in the UV flux from  
 $\eta$ Car itself, the primary rather than its less massive companion,  
 related to $\eta$ Car's recovery from its great eruption.  

Smith et al. (2003b) and van Boekel et al. (2003) showed that the wind of $\eta$ Car today is
latitude dependent with a stronger, faster wind at the poles and a higher ionization and lower
density wind  in the equatorial region.
Smith et al. suggest that more than a century ago, due its loss of several solar masses in the 
great eruption,  
the {\it surface} of $\eta$ Car was rotating slowly within a dense spherical wind. 
The interior, with most of the mass, would have had  the same higher specific angular 
momentum
as before the great eruption. Thus the surface gradually spun up  until its rotation rate was 
sufficient to form the present aspherical wind. This 
transition can be quite rapid, and  the delay time of $\sim$ 100 yrs is reasonable, 
so it may correspond to the
rapid brightening in the early 1940's and the onset  of the high excitation 
emission due to UV photons which may now escape from the lower density equatorial zone. (See Davidson (2005)
for some of the theoretical considerations.)    
In Smith et al.'s suggestion the spectroscopic events are due to equatorial ejections very likely triggered 
by the  approach of the secondary at periastron (3 -- 4 AU) to the massive primary which is already close
to the Eddington limit. The increased density in the equatorial region due to even  a small mass
loss of order of 10$^{-3}$M$_{\odot}$ would be sufficient to temporarily block the UV radiation. 
We note that HST/STIS observations of the two most recent spectroscopic events (1998.0  and 2003.5)
show that they were  not identical and confirm that $\eta$ Car does indeed eject a shell (Davidson et al 2005,
Martin et al 2006a). Furthermore, comparison of the STIS and UVES data (Stahl et al 2005) from the 
2003.5 event show that it was latitude dependent. Changes in the wind at the higher latitudes 
were less dramatic, the wind became denser at the lower latitudes, and consequently, the spectra from the 
equatorial and polar regions were more alike during the spectroscopic minimum, consistent with the
expectations from an equatorial shell ejection.  
 The longer-term observational record  also shows that these spectroscopic events are not identical either in their duration or their spectroscopic characteristics (See Humphreys 1999, Feast et al. 2001).  

In the next section we discuss  the  origin of the He I emission in $\eta$ Car   
and the implications of its absence prior to 1944.   

\section{The He I Emission Problem}    

According to the consensus view since 1998, $\eta$ Car is probably
a binary system with a highly eccentric  5.5-year orbit.  In most discussions the 
hypothetical hot, massive companion star, which has not been
detected {\it per se\/}, serves three purposes:
(1) Its wind collides with the primary wind in order to produce
the observed X-rays, (2) it provides a  basis for the 5.5-yr  period, and (3) it supplies far-UV photons which
indirectly cause the observed \ion{He}{1}, [\ion{Ne}{3}],
and [\ion{Fe}{3}] emission lines.  However, {\it the absence of
such features before 1944 is not easy to explain in models
that invoke a hot secondary star.\/}  This is particularly true
for the broad \ion{He}{1} lines.  In the previous section we discussed the  progressive
changes in the wind of $\eta$ Car. In this section we describe the  
special difficulties concerning the helium emission.

Since the present-day \ion{He}{1} features have broad varying profiles
and the locations of their strongest emission are spatially unresolved in Hubble Space
Telescope (HST) data, they almost certainly originate less than
100 AU from the star.
Being recombination lines, they are produced in zones of ionized He  rather
than neutral He.  (Forbidden high-excitation lines such as [\ion{Ne}{3}]
and [\ion{Fe}{3}] originate in lower-density ionized He  zones farther out,
see Davidson et al. (1995)  and spectra in the HST archives.\footnote{
    http://etacar.umn.edu/ and http://archive.stsci.edu/prepds/etacar/.})
Figure 12 indicates the most relevant locations for ionized He if
a hot, massive companion star is present.  This refers to the
``normal'' situation, not near the time of a spectroscopic event.
Objects \underline{A} and \underline{B} are the two stars,
usually 15--30 AU apart;  \underline{C} (filling most of the figure)
is the dense primary wind;  region \underline{D} is the faster,
lower-density secondary wind;  and \underline{E} is the shocked
region between the winds.  Proposed ionized He  zones include:
\begin{enumerate}
  \item{Region \underline{1}, the hot inner part of the primary wind.
    The observed \ion{He}{1} brightness is difficult to explain
    in a spherical wind model \citep{JH01}, but Smith et al. (2003b) 
    found strong evidence that the wind is latitude-dependent, not spherical.
    The lower-density, hotter equatorial region may produce
    appreciable \ion{He}{1} emission without the
    assistance of a companion star.}
 \item{Region \underline{2}, a photoionized He  zone in the dense
    primary wind, adjoining the shocked region.  It has long been
    recognized that a hot secondary star would create such a zone
    (Davidson 2001a,b) and a qualitative scenario has recently been
    mentioned by Nielsen et al. (2007).   This is probably the most
    popular choice at present.}
 \item{If the secondary star is reasonably normal and less than
    3 million years old (consistent with the massive primary),
    then its wind is not a promising source for $\eta$ Car's
    present-day \ion{He}{1} lines.   The winds of normal
    30--60 $M_{\odot}$ stars at moderate stages of evolution
    are not dense enough to convert more than small fractions of the
    far-UV radiation, and their high wind speeds cause very
    broad line profiles.  Kashi \& Soker (2007), however, have described
    a model wherein they {\it assumed\/} that \ion{He}{1}
    emission arises in the small, dense acceleration zone of
    the secondary wind, location \underline{3} in our sketch.}
 \item{The shocked region between the winds generally contains He$^{++}$ 
    rather than  He$^+$  and has high temperatures, unfavorable for
    recombination emission.  If the
    cooling rate is sufficiently fast, though, some localized gas may
    cool below $10^5$ K before moving far.   The result would be a
    chaotic ensemble of photoionized cloudlets or filaments with
    ionized He  at temperatures of the order of 15000 K, amid the uncooled
    shocked gas hotter than $10^6$ K.   Such condensations would
    be very small and dense due to approximate pressure balance
    within region \underline{E}.  This possibility is indicated by
    label \underline{4} in Fig.12.}
    \item{Smith et al. (2003b) speculated that ionizing photons might come from
    low latitudes in the primary wind's pseudo-photosphere, instead
    of a hot companion star.  If so, then such radiation may ionize
    the low-latitude wind out to large radii.  This idea is not shown
    in Fig.12, because the geometry would be different.}
\end{enumerate}

Any successful model for the present-day \ion{He}{1} features must
account for their weakness or absence before  1944.
Suppose that a hot secondary star ($T_{\mathrm{eff}} > 32000$ K) supplies the helium-ionizing 
photons. Motivated by the nature
of the observed spectroscopic cycle, most authors have favored  a
highly eccentric 5.5-year orbit wherein the apastron and periastron
separations are, respectively, about 30 AU and less than 6 AU.
As we noted in \S{5},  obscuration by circumstellar dust plays almost no role in
the following discussion, since the radiation density within 100 AU
of $\eta$ Car is too high for  dust grains to exist there.

The spectroscopic events now observed at 5.5-year intervals involve,
among other effects, a temporary fading of all emission lines that
originate in the ionized He  regions.   Each spectroscopic event
resembles what one would expect to see if the periastron approach
of star \underline{B} triggers a massive enhancement of the primary
wind at low latitudes.  Helium-ionizing radiation from \underline{B}
can temporarily disappear within the pseudo-photosphere because the
entire configuration is then small and dense, only a few AU across.
Similar suggestions  have been expressed by Zanella et al. (1984),
Davidson (1999), Smith et al. (2003b), and Martin et al. (2006a).  Independent of
whether this general idea is the right explanation for the observed
events, it cannot account for a {\it long-term\/} weakness or absence of
the \ion{He}{1} lines, as we explain in the next paragraph.

The basic difficulty is that hypothetical star \underline{B}
spends most of its time farther than 15 AU from the primary,
at radii where no credible wind can obscure the ionized He  zones.
The primary wind may have been denser before 1944 than it is
today, particularly at low latitudes as emphasized by \citep{NS03b}.
High densities would not have prevented star \underline{B} from
ionizing helium in region \underline{2} of Fig. 12, and the
standard efficiency for converting far-UV photons to helium
recombination emission depends only weakly on density \citep{of06}.
Suppose, therefore, that we try to hide the \ion{He}{1} recombination
lines by continuum absorption and scattering in the dense wind.
Region \underline{2} in Fig. 12 is usually at distances
of 15 to 35 AU from the primary.  In order to produce a
Thomson-scattering optical depth of unity outside
$r  \sim  25$ AU, a spherical model with a wind speed of
500 km s$^{-1}$ would need a mass-loss rate of about
0.01 $M_{\odot}$ yr$^{-1}$, ten times the present-day value.
But this amount of scattering would merely distort the emission line
profiles.  In order to suppress or conceal them, either the optical
depth for scattering must be very large, or, more likely, absorption
must be strong enough that the ``thermalization optical
depth'' is substantial, $\sqrt{3 {\tau}_{\mathrm {abs}} {\tau}_{\mathrm {tot}}}$
$\gtrsim$ 1, around the ionized He  region.  This would automatically
create a pseudo-photospheric radius of the order of 25 AU, implying an
effective temperature less than 4500 K for the entire configuration.
Such a model is unlikely for several reasons:
\begin{itemize}
   \item{It would be even larger and cooler than the state seen in
      the second eruption, e.g.\ around 1893.  In that case one does not
      expect to see the emission-line spectrum that was observed in
      1900--1941, described elsewhere in this paper and qualitatively
      resembling the present-day forest of \ion{Fe}{2}, [\ion{Fe}{2}],
      and \ion{Ni}{2} emission lines.}
\item{Relevant opacities decline rapidly at temperatures below
7000 K;  this is why supernova explosions and major LBV eruptions
have pseudo-photospheres around 7000 K (Davidson 1987).
Therefore, if the hypothetical opaque gas was cooler
than 5000 K as one would expect from the size scale
and $\eta$ Car's luminosity, then it must have been several
orders of magnitude denser than present-day conditions, in order to have sufficient optical depth.
This would imply a tremendous 1900--1940 mass-loss rate whose
ejecta should have been obvious today.\footnote{
     A detailed quantitative discussion of this point would be
     beyond the scope of this paper because the theoretical
     situation is complex.  However, our comments are
     illustrated by giant LBV eruptions, see
     Humphreys \& Davidson (1994).  The ``Weigelt blobs''
     located several hundred AU from $\eta$ Car are not
     massive enough to be the 1900--1940 ejecta in this
     type of model, see Davidson et al.\ (1995). }
Larger opacities would have required higher temperatures,
6000--10000 K at locations $r > 15$ AU.  Opaque gas with
$T > 6000$ K  obviously did not surround
the star at  $r > 15$ AU, because the emergent luminosity would have
exceeded $10^7  L_{\odot}$, more than twice the present-day
total for $\eta$ Car.  Thus we must imagine a localized
absorbing region,  but its characteristic diameter had to
be at least 10 AU in order to hide zone 2 in Fig.\ 12.
It would have radiated more than $10^6 L_{\odot}$, while
covering only a small fraction of the solid angle around
$\eta$ Car -- in other words, one must explain how enough
energy was channeled through such a region to
heat it. }
\item{With temperatures of 4000--10000 K as mentioned above,
the hypothetical opaque gas would have radiated mainly
at blue-to-red wavelengths, not in the UV as $\eta$ Car does
today.  Since it would have been intrinsically very bright at
photographic and visual wavelengths, at least 5 magnitudes
of circumstellar and interstellar extinction would have
been required in order to explain the 8th-magnitude
appearance seen before 1940.  A strongly obscured, relatively
cool radiator would have appeared very red, $B - V > 2$,
contrary to the observations (including the objective
prism spectra). }
\end{itemize}

The same considerations apply to region 4 in Fig.\ 12.
A resourceful theorist might overcome these difficulties
by carefully adjusting the geometry and other parameters,
but {\it no quantitative model of this type has yet been
proposed.\/}

As we remarked earlier, the fast secondary wind is an unlikely
source for the observed \ion{He}{1} emission.  Normal O-type stars
do not produce \ion{He}{1} lines with strengths and profiles
comparable to those now seen in $\eta$ Car.  Kashi and Soker (2007)
nevertheless propose that the region of interest is in the acceleration
zone of the secondary wind, a locale with radius less than 0.2 AU.
Maybe the primary star and its wind somehow alter the physical
conditions there.  If so, then those conditions may have been
different before 1940;  the arguments listed above do not apply
to a source region smaller than several AU.  One obvious objection
to this scenario, however, is that the {\it inner\/} wind of a
massive O-type star should be physically robust in this context.
The acceleration zone's ram pressure, gas density, radiation density,
and ambient temperature considerably exceed the values found in
nearby regions of the primary wind at most phases of the orbital
cycle.  (If they do not, then a colliding-wind model has difficulty
explaining the X-rays, see Pittard \& Corcoran 2002.)  A substantial
perturbation of the inner secondary wind, needed in order to produce
strong \ion{He}{1} emission there, would evidently require some
special, rather surprising effect.    Moreover, if the secondary
wind accounts for the present-day \ion{He}{1} lines, then one must
explain why zones 2 and 4 in Fig.\ 12 -- the large ``obvious''
regions for \ion{He}{1} emission -- are ineffective.

Alternatively one might invoke some unusual type of secondary
star or secondary wind;  but such models are unappealing because
they require two peculiar objects together.

 In summary, one cannot explain the pre-1944 absence of \ion{He}{1}
lines merely by supposing that  ``the ionized He region was obscured''
or ``outflowing material engulfed the companion star.''
{\it If a hot secondary star provides most of the helium-ionizing
photons, then some additional effect is required in order to explain
the pre-1944 weakness of the helium features.\/}

One aspect of helium recombination may help.  In the following
account, we employ atomic parameters quoted by Osterbrock \& Ferland (2006) 
and parameters for $\eta$ Car consistent with various
observations \citep{dh97}.  This is an
order-of-magnitude outline in order to put the problem
in context; we plan a future, more detailed discussion of the
present-day \ion{He}{1} intensities and velocity profiles. The
threshold for photoionization of helium in its ground level
1s$^2$ $^1$S is $\epsilon \approx 24.6$ eV or
$\lambda \approx 504$ {\AA}.  Most helium recombination events 
 create atoms in excited triplet states, which then
decay in one or more steps to the metastable 1s2s $^3$S level about 19.8 eV
above the ground state. For
densities of interest here, subsequent decay to the singlet ground
level is enabled mainly by collisional transitions
from 1s2s $^3$S to the 1s2s and 1s2p singlet states.  If this were the
whole story, then the average lifetime of the He$^0$ 1s2s $^3$S level
would be of the order of 0.1 s at radius $r \sim 25$ AU in
$\eta$ Car's present-day wind.

However, that level can be ionized by near-UV photons
having $\epsilon > 4.8$ eV and $\lambda < 2600$ {\AA}.  Given the
present-day spectrum of $\eta$ Car, this should usually
occur in a time less than 0.001 s at the location mentioned above --
much faster than transitions to singlet states.   In an ionized 
zone near $\eta$ Car, therefore, most triplet recombination events
do not create ground-level helium atoms.   If we count only
singlet recombination events, the effective recombination coefficient
for He$^+$ $\rightarrow$ ground-level He$^0$ is thereby reduced by
a factor of roughly 4 compared to standard nebular conditions.
The volume extent of an ionized He  zone is consequently larger than
one would estimate from a Zanstra calculation based on photons above
24.6 eV and total recombination coefficients.  Meanwhile the triplet
recombination events {\it do\/} lead to decays which produce emission
lines, including $\lambda$4471.5.  Result:  {\it For a given amount
of radiation above 24.6 eV, $\eta$ Car's present-day wind produces
far more helium recombination emission than one would naively expect.\/}

Conceivably we can explain the pre-1944 \ion{He}{1} emission-line
deficit by supposing that the standard textbook recombination
coefficient was valid at that time, thereby reducing the extent
of the ionized He  zones.  In other words, maybe the 1s2s $^3$S
re-ionization effect described above occurs today but was ineffective
before the 1940's.  Here are two possible reasons:
\begin{itemize}
  \item{If $\eta$ Car's mass-loss rate was quite high at that time,
     then the pseudo-photospheric surface in the opaque primary wind
     would have been larger and cooler than it is today.  Maybe it
     produced very little radiation capable of ionizing
     He$^0$ 1s2s $^3$S ($\epsilon > 4.8$ eV, $\lambda < 2600$ {\AA}). }
  \item{Higher densities in the ionized region would have increased
     the rate for collisional transitions from 1s2s $^3$S to singlet
     states.   This would be especially true if the present-day
     ionized He  zone is region \underline{2} of Fig. 12, but before
     1944 it was in region \underline{4} because cooling
     was faster then. As noted above, small condensations of cooled
     material can have very high densities within the shocked region.}
\end{itemize}
These suggestions do not solve the problem, though, since at least three
objections must be overcome:
\begin{itemize}
  \item{Independent of the primary star, the hypothetical
     companion star can also ionize He$^0$ 1s2s $^3$S.  For every type
     of proposed object, photons above 4.8 eV far outnumber those
     above 24.6 eV.}
  \item{The estimated ratio of rates for depopulating the 1s2s $^3$S level,
     \begin{displaymath}
     r({\mathrm {photoionization}}) / r({\mathrm {collisional}}),
     \end{displaymath}
     must be decreased by a large factor, on the order of 300,
     in order to make triplet recombination events effective in
     populating the ground level.}
  \item{The effect that we have described can change the amount
     of ionized He  by a factor of about 4 but not much more.  As
     discussed in \S{4}  above, the \ion{He}{1} $\lambda$4471 line 
     may be present in the 1938 spectrum with at most one-fourth its 
     strength in 1944-51 
     (relative to other emission lines), but the data actually support a
     lower value.}
\end{itemize}

Now suppose we consider a different possibility, that $\eta$ Car's
present-day \ion{He}{1} lines are excited by the primary star and
have nothing to do with a companion -- i.e., cases
1 and 5 in our list of He$^+$ locations.  Then the pre-1944 spectra
present no serious difficulty
concerning helium;  a modest increase in the wind density,
or a change in latitude structure as suggested by
Smith et al (2003b) and Davidson (2005), would suppress helium-ionizing photons
and emergent \ion{He}{1} emission lines (see \S{5.4}).  Since this would
occur in the inner wind, $r < 6$ AU, the difficulties
noted above would not apply.  In a model of this type the
high excitation lines in general, e.g. [\ion{Ne}{3}] and
[\ion{Fe}{3}],  may be unrelated to any hypothetical  secondary star.

In summary:  Given that \ion{He}{1} emission lines have been quite
obvious in $\eta$ Car since the mid-1940's, {\it their weakness
or absence before that time is difficult to explain  in models
that employ a companion star to supply the far-UV photons.\/}
We do not claim that this argument disproves the concept of
a hot secondary star, but it does imply that the popular binary
scenario requires additional justification which no one has
yet offered.  So far, many qualitative ideas have been proposed
for $\eta$ Car's spectroscopic behavior but there are no realistic, quantitative
{\it models.\/}

\section{Conclusions} 

Our measurement and analysis of the Harvard objective prism spectra
of $\eta$ Car from 1902 to 1941, has shown that He I 4471{\AA} emission may be marginally present, 
but  no stronger than about  one-fourth of  its relative intensity  as estimated by Gaviola in spectra
from 1944-51.  Our results are equally consistent with no contribution from the He I $\lambda$4471 line. 
The [Fe III]  high excitation lines are either absent or very weak in these spectra.
The star was thus always in the ``low excitation'' state and there were no 
apparent spectroscopic events from the time of the second eruption in the 1890's to 1944.     

Obscuration of the proposed companion star by either dust or an optically thick
wind cannot explain the lack of He I emission prior to 1944. 
We discussed the possible sources
of He I emission in the $\eta$ Car system and demonstrated  that it is very difficult to explain
its absence if the UV flux comes from the proposed
hot secondary star. Alternatively, we suggest that the equatorial wind of the primary 
may be the source of the UV flux due to
$\eta$ Car's post-eruption rotational spin-up and the onset of its present
bipolar wind, coincident with its brightening between 1940 and 1952 (Smith et al. 2003b). 

This work has also demonstrated that with careful processing and calibration these historical 
spectra are quite useful especially for objects of astrophysical interest.  

\acknowledgments  
RMH is especially grateful to Alison Doane, Tracy McGinnis and Martha Hazen at CfA for their assistance
with the plate collection and for digitizing several of the plates.  
We also thank John Hillier for sharing his 1986 spectrum of $\eta$ Car with us, and
Nolan Walborn for comments on his identifications in Le Sueur's 1870 spectrum.

\appendix

\section{The Harvard Objective Prism Plates of Eta Car}

For future reference, we provide  a chronological list of all the Harvard objective prism plates with
spectra of $\eta$ Car in Table A1 that we examined including comments on the spectra, the image quality, and in
some cases whether  the plate was missing or broken.


\newpage 

\begin{figure} 
\figurenum{1}
\epsscale{1.0}
\plotone{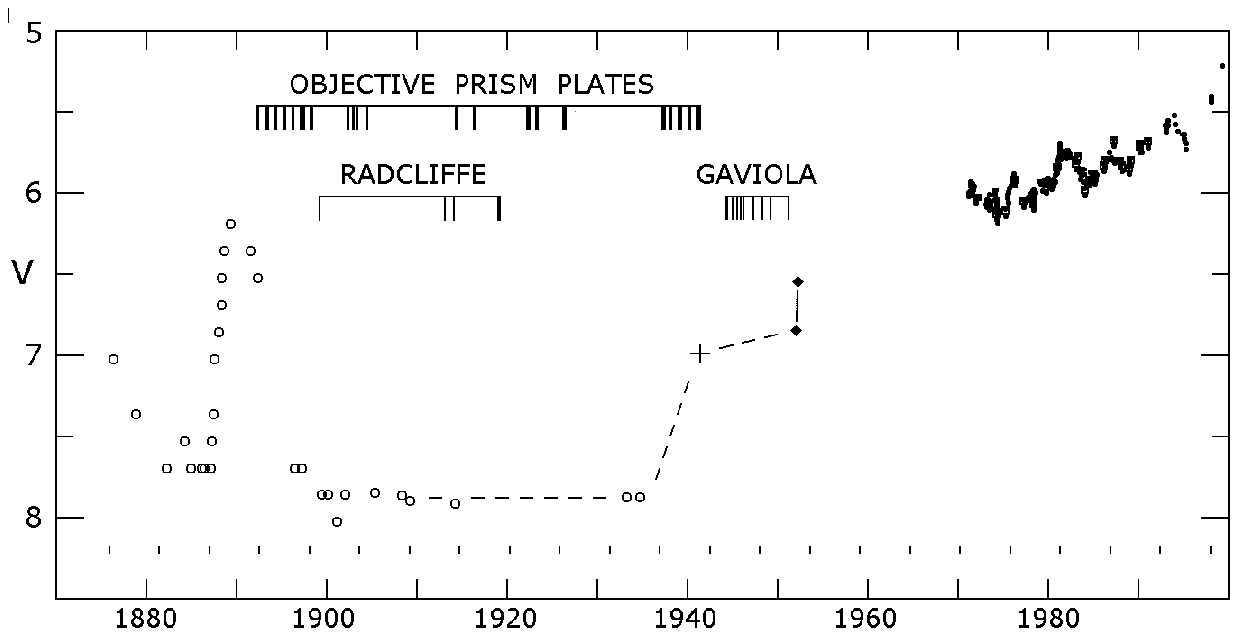}
\caption {The light curve of $\eta$ Car from 1870 to the present showing the times
of the early spectra. The small tick marks near  the bottom indicate the times of spectroscopic events 
($t = 2003.5  - 5.54N$). Events before 1948 were not observed and the period is not guaranteed to be constant.}
\end{figure}

\begin{figure}
\figurenum{2}
\epsscale{0.9}
\plotone{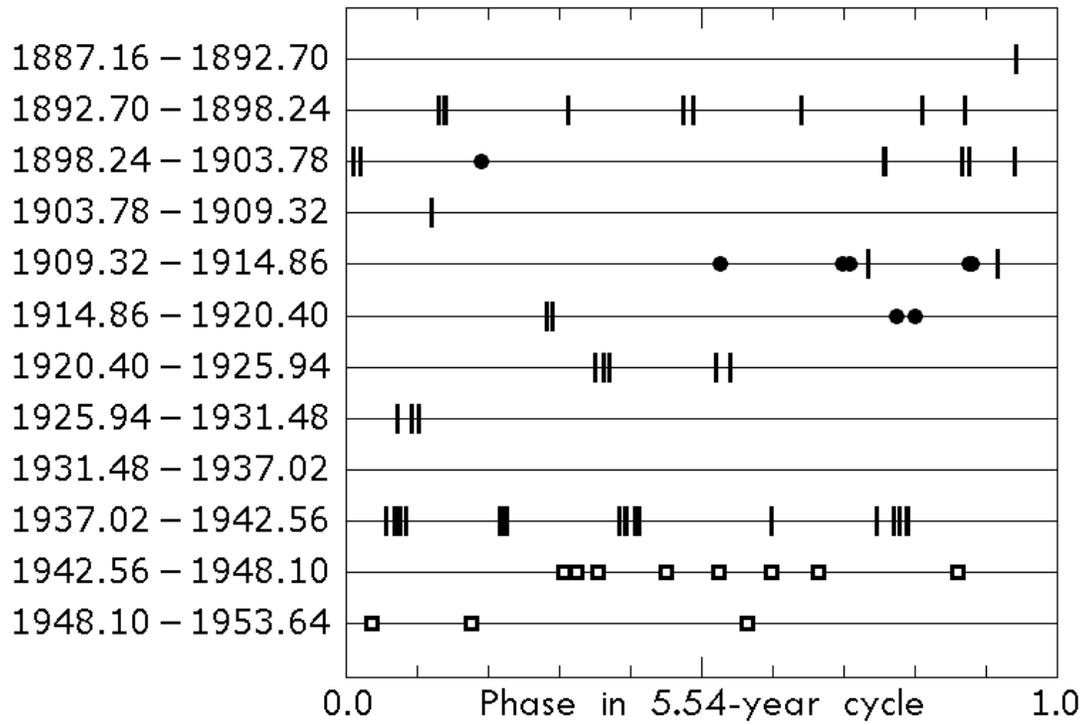}
\caption {The phase of the early spectra in the 5.5 yr. cycle. The vertical marks are the Harvard objective prism spectra, the dots are  Radcliffe spectra (Feast et al 2001), and the boxes are Gaviola's (1953) spectra.}
\end{figure}

\begin{figure}
\figurenum{3}
\epsscale{0.7}
\plotone{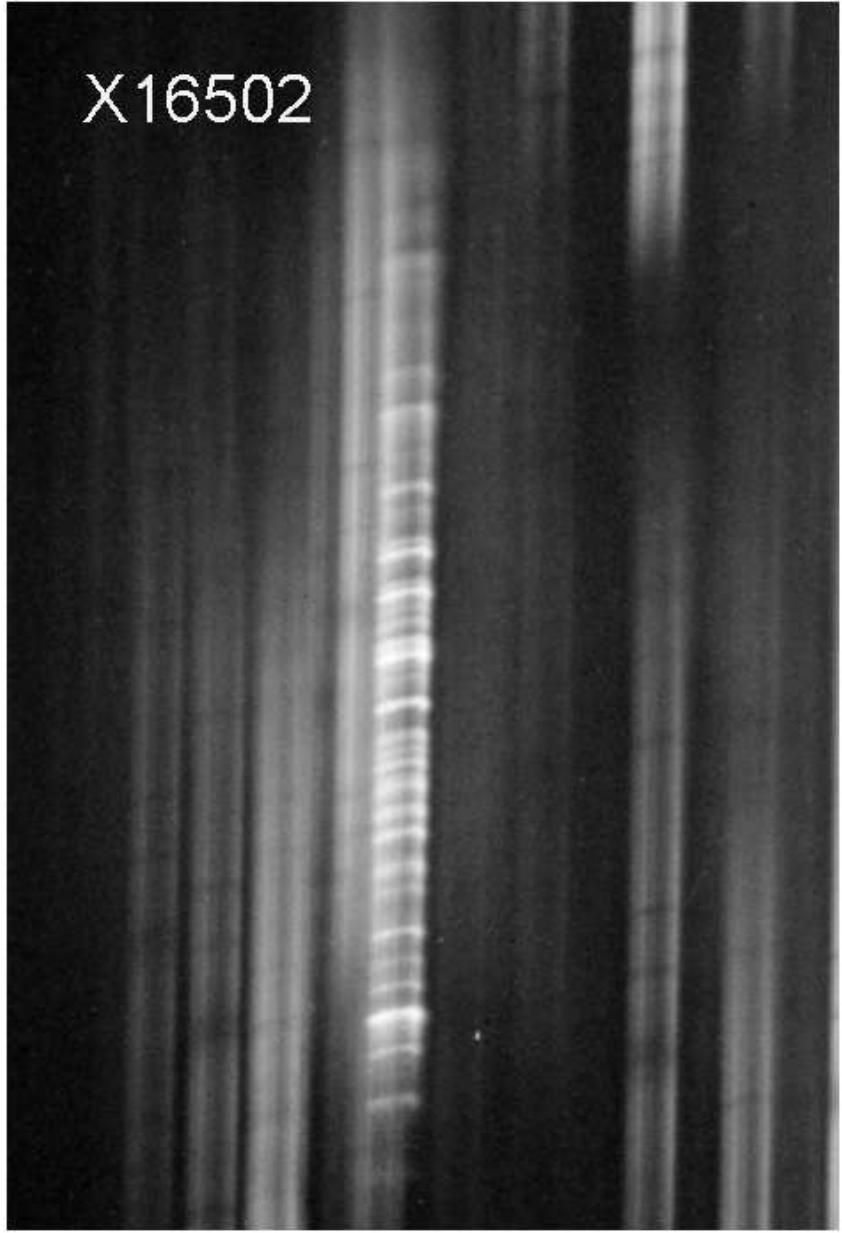}
\caption {The digitized section of objective prism plate X16502; only a small fraction of the plate was digitized. Since this is a long exposure,  the spectra of other
nearby stars are conspicuous.  The breadth of the spectrum perpendicular
to the dispersion, caused by drifting the telescope, was slightly
less than 1{\arcmin} in this example. Note that the dispersion direction is not
exactly
vertical in the digitized file.}
\end{figure}

\begin{figure}
\figurenum{4}
\epsscale{1.0}
\plotone{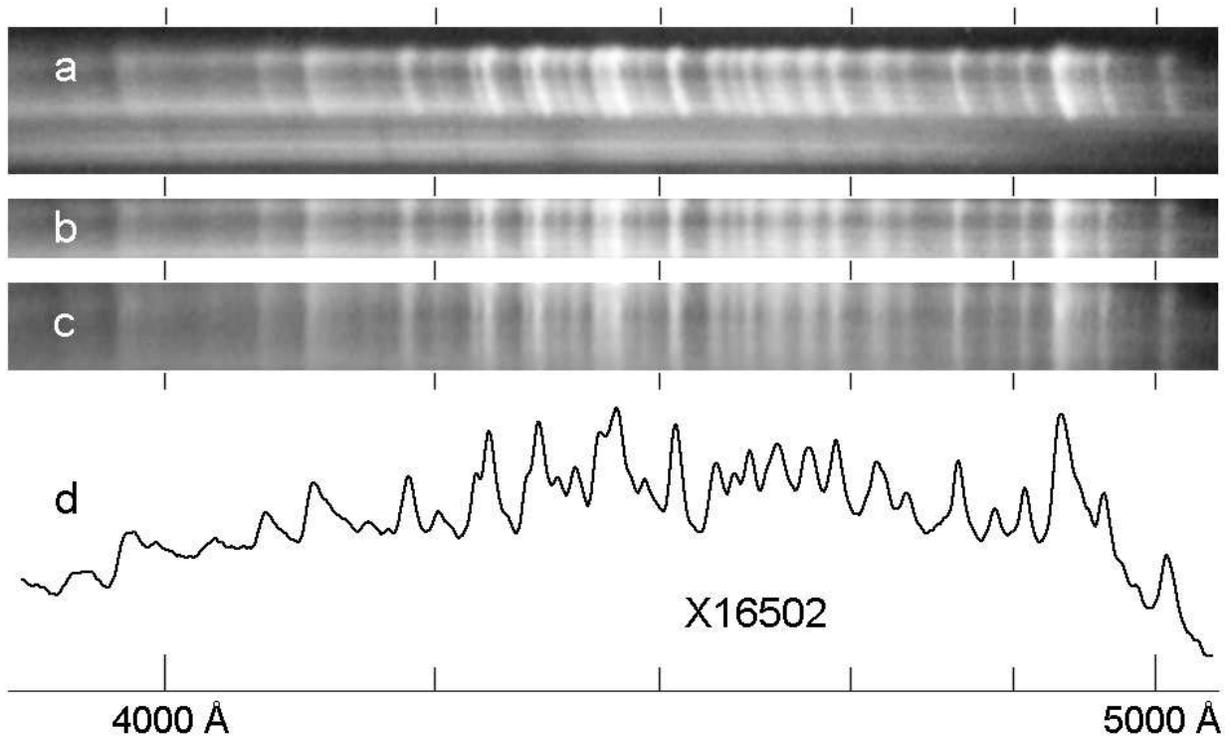}
\caption {X16502.  (a) The original digitized
scan, like Fig.\ 3 but rotated so the dispersion direction runs
exactly along rows of pixels.   (b) Corrected for the curvature and other guiding
problems as explained in the text.  (c) Broadened
perpendicular to the dispersion, and renormalized so that every
row has the same average level.  (d) The tracing of the spectrum,
extracted by a procedure described in the text.  Since the data
are photographic with largely unknown parameters, the vertical
scale is not linear.  The tick marks in (a), (b), and (c) correspond
to the same wavelength scale as (d).  Plate X16502 was an unusually
long exposure whose spectrum of $\eta$ Car is of lower quality
than most of the spectra used  in our analysis.
It was chosen here because it provides a clear example of our processing steps.}\end{figure}

\begin{figure}
\figurenum{5}
\epsscale{1.0}
\plotone{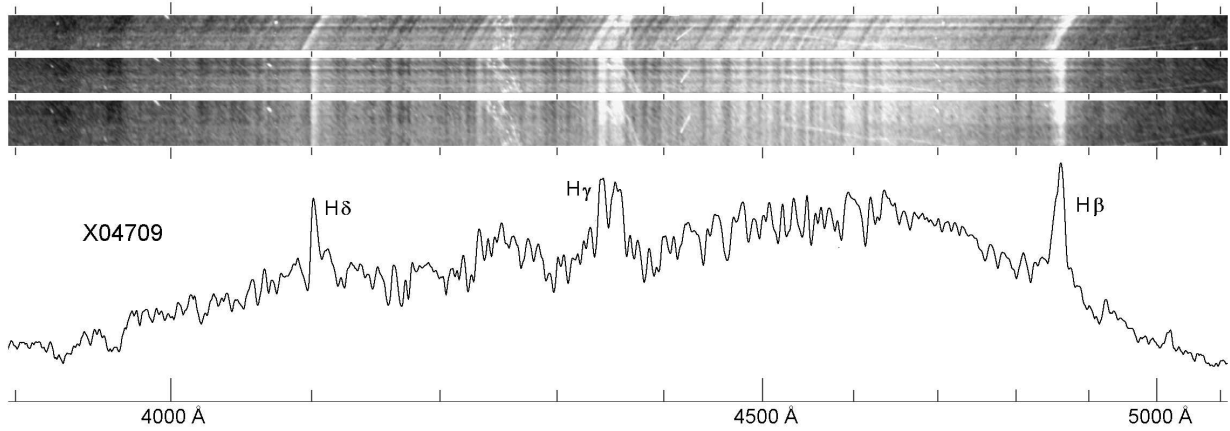}
\caption{X4079, the famous spectrum from 1893 during $\eta$ Car's second eruption. The format
is the same as figure 4. The upper spectrum  indicates that the guiding was unusually steady
for this plate.
Most of the irregularities in the tracing are real spectral features,
not noise. The plate has serious scratches in the emulsion near 4250 and 4360{\AA}, but
our processing techniques reduced their effects in the tracing.}
\end{figure}

\begin{figure}
\figurenum{6}
\epsscale{1.0}
\plotone{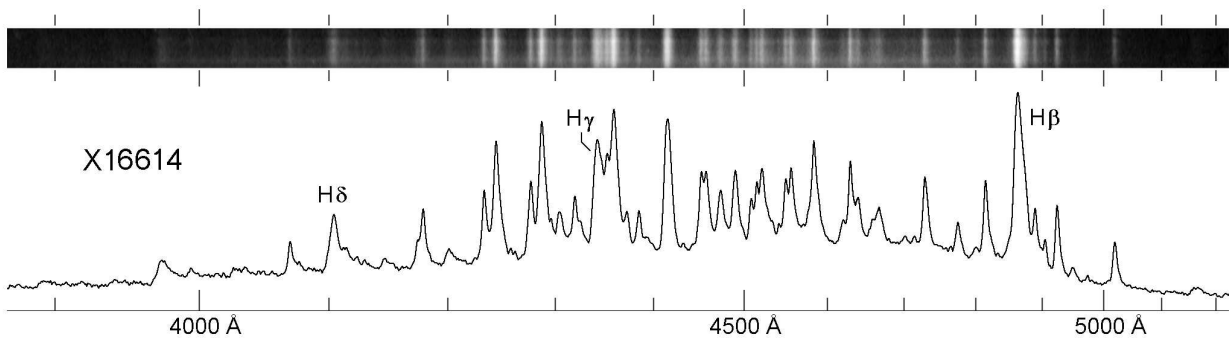}
\caption{X16614(1938) is one of the best spectra in the objective prism series.
The top spectrum shows  the  digitized scan rectified by our technique and the
extracted tracing is shown below it.}
\end{figure}

\begin{figure}
\figurenum{7}
\epsscale{1.0}
\plotone{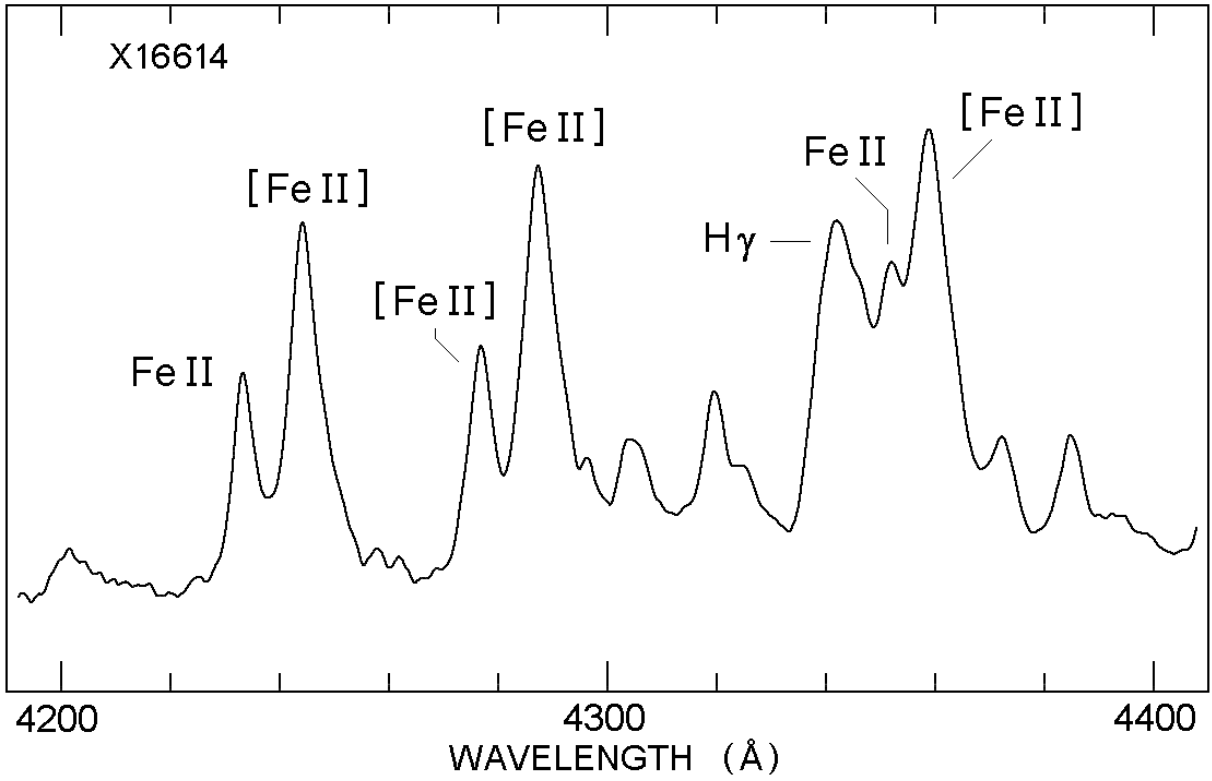}
\caption{The spectral region around H$\gamma$ in X16614. Caveat: The vertical scale is non-linear becasue it represents photographic densities.}
\end{figure}

\begin{figure}
\figurenum{8}
\epsscale{1.0}
\plotone{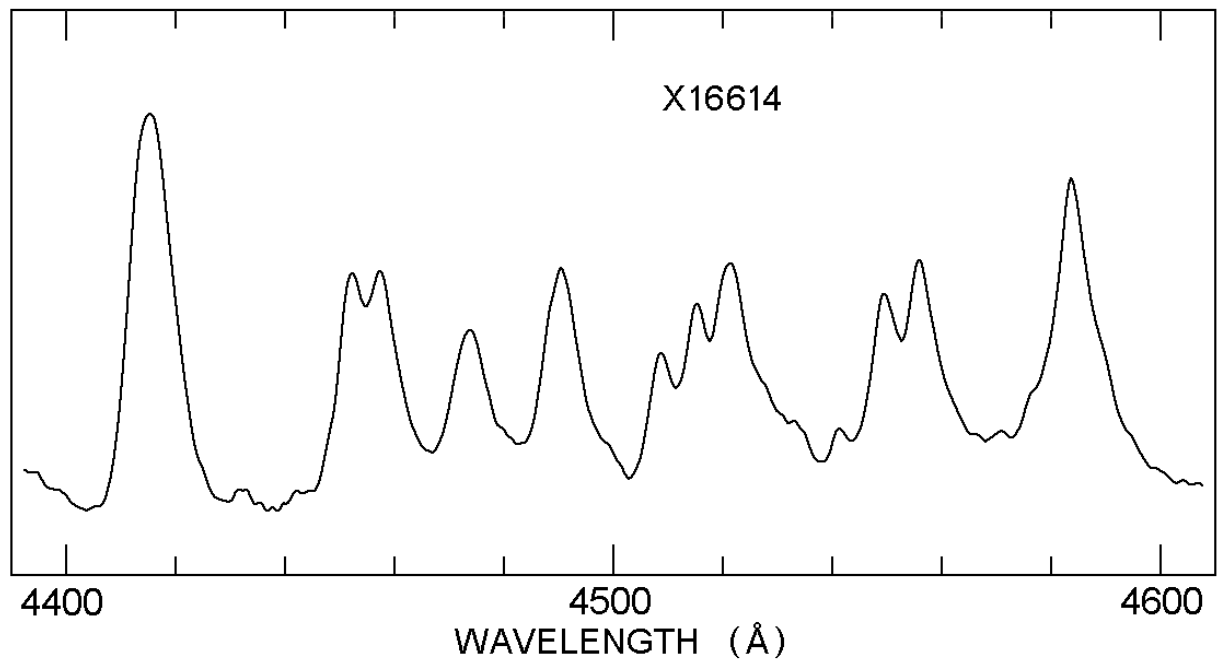}
\caption{The main features shown here are all lines or blends of
[Fe II] and Fe II. The region near 4470{\AA}, where
He I emission might appear, is shown in more detail
in Fig. 9 }
\end{figure}

\begin{figure}
\figurenum{9}
\epsscale{1.0}
\plotone{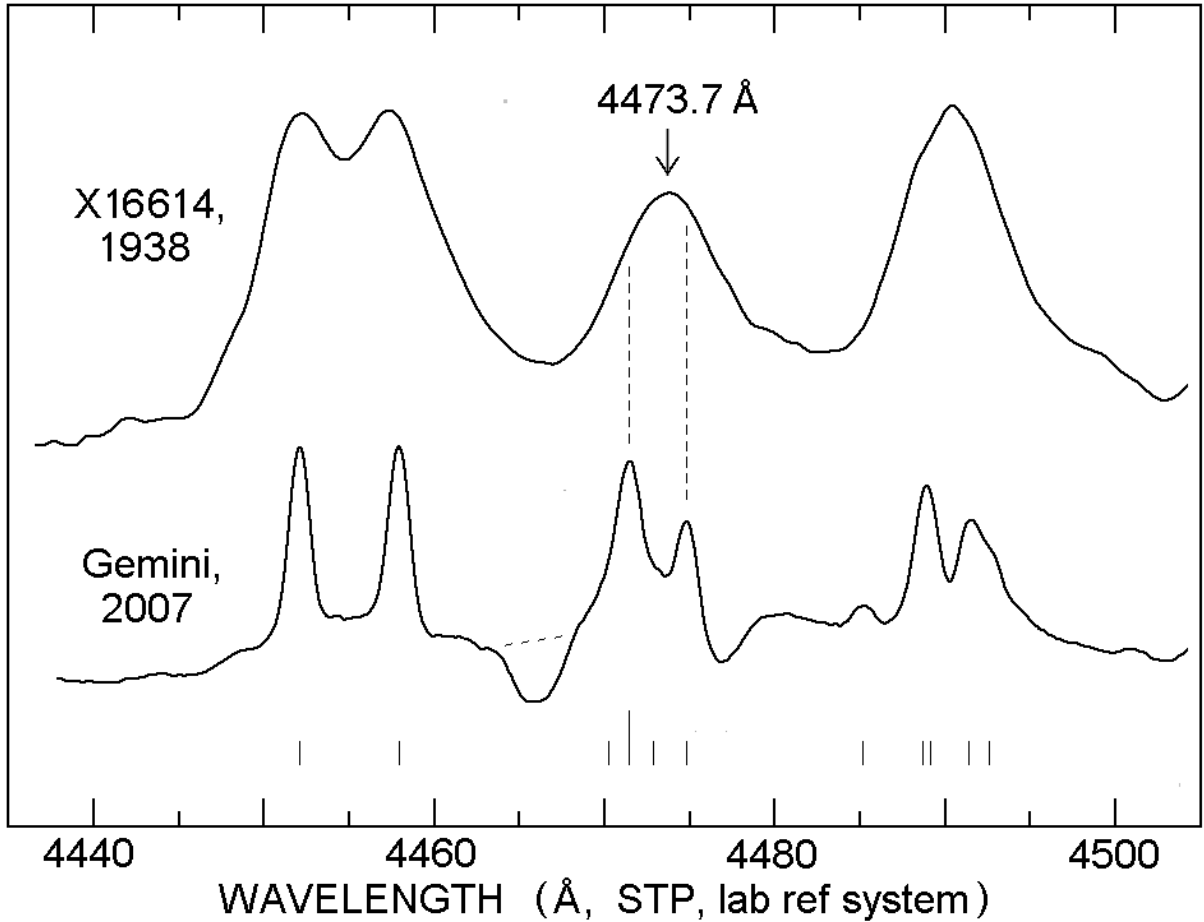}
\caption{The upper spectrum shows the expanded spectrum of  X16614 in the wavelength region of He I $\lambda$4471.5.
The vertical dashed lines mark the expected positions of He I $\lambda$4471.5 and the [Fe II] line at 4474.9{\AA}.
The spectrum below it is the  Gemini-S spectrum from June 2007. The narrow vertical lines mark the laboratory wavelengths of identified Fe II and [Fe II] lines
in Aller \& Dunham's Coud$\acute{e}$ spectra. The dashed horizontal line in the
Gemini spectrum marks the dip due to He I P Cygni absorption.}
\end{figure}

\begin{figure}
\figurenum{10}
\epsscale{1.0}
\plotone{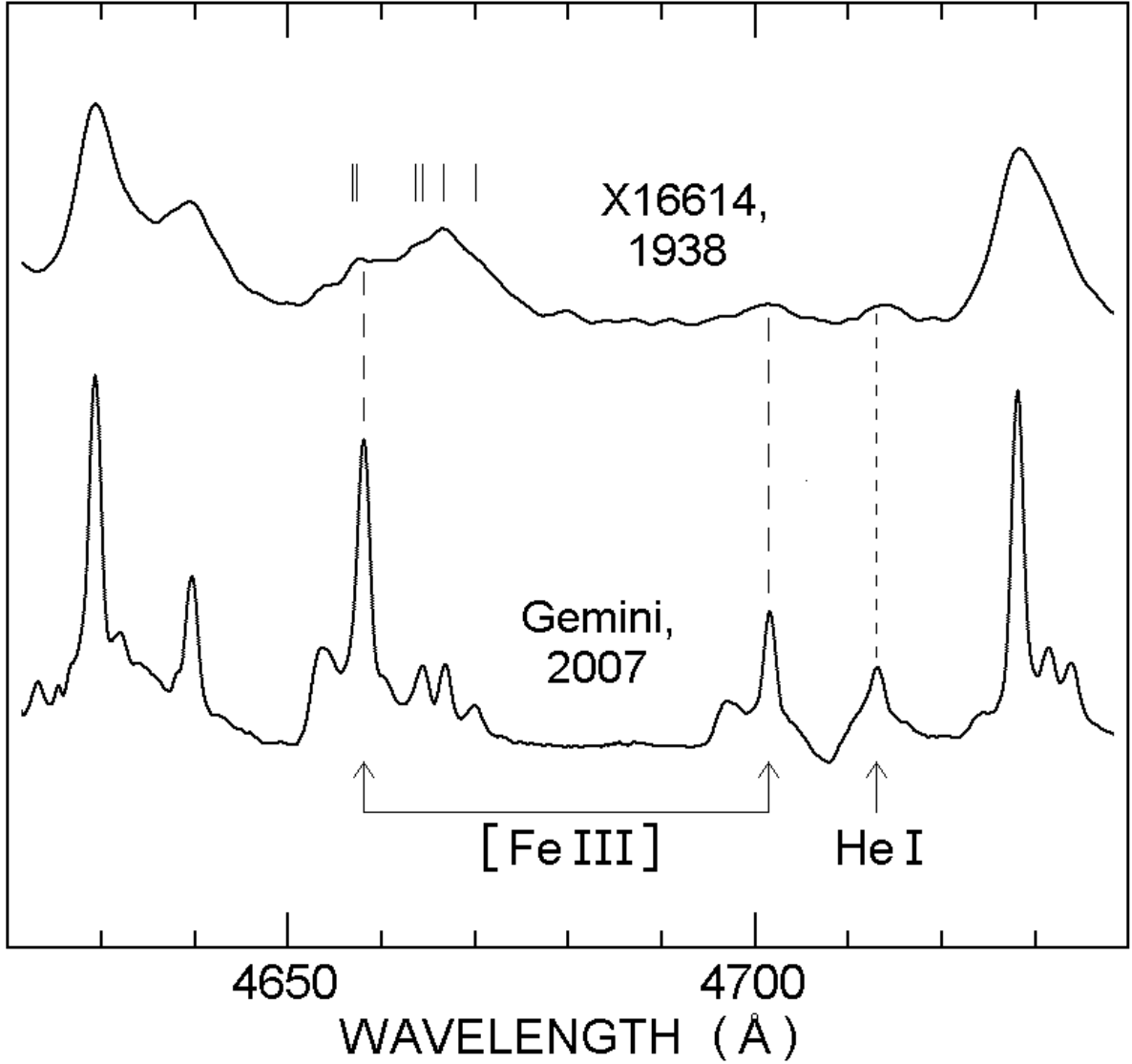}
\caption{The expanded spectrum of X16614 in the region of the [Fe III] $\lambda$4658.1
emission line. The tick marks at the top represent the low-excitation lines mentioned
in the text, and  several weak lines of Fe II and Cr II may also be contributingto the profile. The small bump in the X16614 spectrum at $\lambda$4701 may be due to [Fe III], but the other weak feature to its red does not  match the He I line in
wavelength.}
\end{figure}

\begin{figure}
\figurenum{11}
\epsscale{1.0}
\plotone{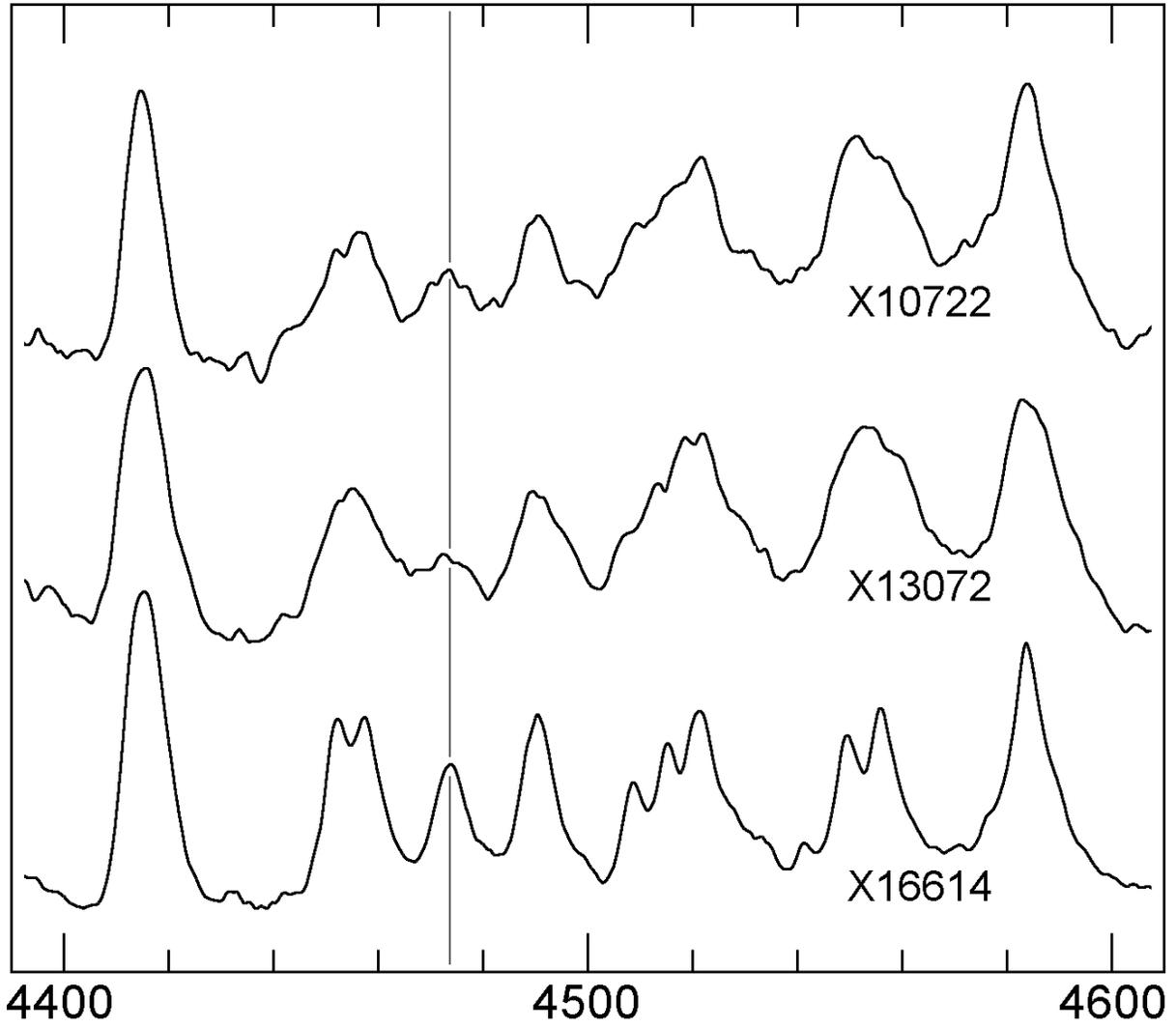}
\caption{Two early spectra, X10722 (1903) and X13072 (1916) together with X16614in the region of the $\lambda$4473 emision feature. The vertical line marks the
wavelength 4473.7{\AA}.}
\end{figure}

\begin{figure}
\figurenum{12}
\epsscale{0.8}
\plotone{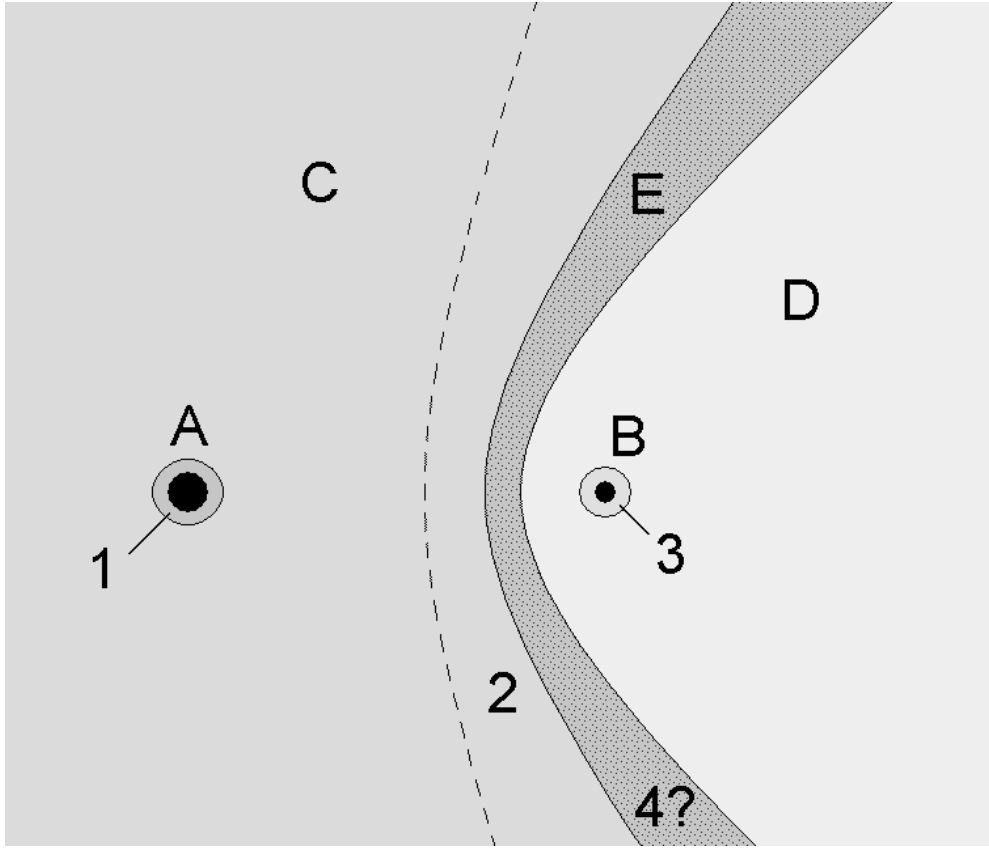}
\caption{Schematic arrangement of distinct zones in a colliding-wind model
with a hot secondary star.  \underline{A} and \underline{B}:
The two stars.  \underline{C}:  The dense primary wind.
\underline{D}:  The faster, less dense secondary wind.
\underline{E}:  Hot shocked gas.
\underline{1}:  The inner wind of the primary star may produce
\ion{He}{1} emission lines.
\underline{2}   He$^+$ zone in the primary wind, photoionized
  by the secondary star.
\underline{3}   Acceleration zone of the secondary wind, see text.
\underline{4}   Small, dense cooled structures might exist within
  the shocked region, see text.}
\end{figure}


\newpage

\begin{deluxetable}{lllll}
\tablecaption{Harvard Objective Prism Plates}
\tablewidth{0pt}
\tablehead{
\colhead{Series}  & \colhead{Telescope} &  \colhead{Plate Scale}  &  \colhead{Dates}  &
 \colhead{Location}\\
}
\startdata
A Series  &  24-inch Bruce Doublet &  60$\arcsec$ mm$^{-1}$   &  1896 -- 1926 & Arequipa, Peru \\
 \nodata  & \nodata       &    \nodata  &  1929 -- 1950 & Bloemfontein, South Africa \\
B Series  &  8-inch Bache Doublet &  179$\arcsec$ mm$^{-1}$  &  1891 -- 1923  &  Arequipa, Peru \\
 \nodata  & \nodata       &    \nodata  &  1930 -- 1954 & Bloemfontein, South Africa \\
X Series  & 13-inch Boyden Refractor &  42.4$\arcsec$ mm$^{-1}$  &  1891 -- 1926  & Arequipa, Peru \\
 \nodata  & \nodata       &    \nodata  & 1930 -- 1951  & Bloemfontein, South Africa \\
\enddata
\end{deluxetable}

\begin{deluxetable}{lcccc}
\tablecaption{High Excitation Relative Line Strengths}
\tablewidth{0pt}
\tablehead{
\colhead{Source} & \colhead{Date\tablenotemark{a}} & \colhead{Phase} &  \colhead{He I 4471.5/[Fe II] 4474.9}  &  \colhead{[Fe III] 4658/[Fe II] 4639.7} 
}
\startdata
Gaviola (1953)  &   1944 - 1951& ---   &    1  &  0.8 \\
Thackeray(1953) &    1951      & .58   &   1.2  & 0.7 \\
Aller \& Dunham(1966) &  1961   & .39  &   1.2  & 1.0 \\
Hillier \& Allen\tablenotemark{b} & 1986 & .88  &   1.1  & 1.2 \\
Zethson (2001)           & 1999 & .21  &   1.2  & 1.4 \\
Gemini\tablenotemark{c}  & 2007 & .71  &   1.2  & 1.4 \\  
 
\enddata
\tablenotetext{a}{The date of the observation(s)}
\tablenotetext{b}{Spectrum courtesy of John Hillier}
\tablenotetext{c}{From a spectrum obtained with GMOS on Gemini-S, June 2007} 
\end{deluxetable}


\begin{deluxetable}{lllccl}
\tablenum{A1}
\tabletypesize{\scriptsize}
\tablecaption{The Harvard Objective Prism Plates of Eta Car}
\tablewidth{0pt}
\tablehead{
\colhead{Plate Number} &  \colhead{Calendar Date}  & \colhead{Julian Date} & \colhead{Exposure Time} & \colhead{Phase}  & \colhead{Comment}\\
& & &  \colhead{minutes} &  &  \\ 
}
\startdata
X4005    &   5-15-1892  & 2412234  & 126 & .94 &  missing \\
X4044    &   5-20-1892  & 2412239  & 60 & .94  & very weak exposure\tablenotemark{a}\\      
X4709\tablenotemark{b}    &   6-2-1893  &  2412617  & 37 & .13  & see text \\
X4725    &   6-6-1893   & 2412621  & 77 & .13   & narrow, underexposed\\
X4801    &   6-17-1893  & 2412632  & 70 & .13   & weak, on plate edge\\
X4846    &   6-23-1893  & 2412638  & 60 & .14  & weak, on plate edge\\
X4868    &   6-26-1893  & 2412641  & -- & .14   & poor plate, all underexposed\\
X4869\tablenotemark{b}    &   6-26-1893  & 2412641  & 60 & .14  & weak exposure\tablenotemark{a}\\
X5568\tablenotemark{b}    &   6-8-1894   & 2412988  & 60 & .31  & weaker than X4869\tablenotemark{c}\\
X6358    &   4-17-1895   & 2413299  & -- & .47   & underexposed\\
X6480\tablenotemark{b}    &   5-2-1895    & 2413314  & 60 &  .48 & no P Cyg profiles on emission lines\tablenotemark{d}\\  
X6592\tablenotemark{b}    &   5-25-1895   & 2413337  & 64 &  .49 & no absorption lines, H and Fe emission\\
X7415    &  4-8-1896    &    2413658     &  60 & .64 & $\eta$ Car not visible, but should be there!\\
A2224    &  3-12-1897   &    2413997     &  77 & .81 & H$\beta$ emission\\  
B19639   &  6-29-1897   & 2414106        &  121 & .87 &  low disprersion, H$\beta$ emission \\
B21154   &  4-11-1898   & 2414392        & 107  & .01 & low disprersion, H$\beta$ emission \\
B21379   &  5-12-1898   & 2414423        &  158 & .02 &  low disprersion, H$\beta$ emission \\
A5787    &  4-8-1902    & 2415849        & 60 & .73  & looks like all later spectra\tablenotemark{e}\\ 
X10482\tablenotemark{b} & 6-5-1902 & 2415906 & 60 & .76  &  -- \\
X10494   & 6-7-1902     & 2415908 & 58 & .76 & --\\
X10505   & 6-10-1902    & 2415911 & 60 & .76 & --\\
X10513\tablenotemark{b} & 6-11-1902 &  2415912 &  48 & .76 & -- \\
X10644   & 1-16-1903    & 2416131   & 63 & .87 & plate missing\\
X10649   & 1-17-1903    & 2416132   & 64 & ,87 & plate broken\\
X10658   & 1-20-1903    & 2416135   & 57 & .87 & plate missing\\
X10667\tablenotemark{b} & 2-3-1903  & 2416149   &  60 &  .88 &  --  \\
X10703   & 5-13-1903    & 2416248   & 91 & .93 & blurred \\
X10705   & 5-14-1903    & 2416249   & 63 & .93  & bad focus, plate broken \\
X10713   & 5-15-1903    & 2416250   & 62 &  .93 &  poor guiding \\
X10722\tablenotemark{b} & 6-12-1903 & 2416278 & 65 & .94 &  -- \\  
X10855\tablenotemark{b} & 6-12-1904 & 2416644 & 53 & .12 & --\\
X12999   & 5-23-1913    & 2419901  & 112 & .74  & --\\
X13003   & 6-5-1913     & 2419914  & 87 & .74  & missing \\
X13009   & 6-30-1913    & 2419939  &  -- & .75  & weak exposure, $\eta$ Car not visible\\
X13033   & 5-25-1914    & 2420278  & 60 & .92  & --  \\
X13034   & 5-25-1914    & 2420278  & 60 & .92  & --  \\   
X13035   & 5-26-1914    & 2420279  & 60 & .92  & -- \\
X13059   & 6-3-1916     & 2421018  & 62 & .28  & $\eta$ Car not on plate \\
X13062\tablenotemark{b} & 6-7-1916 & 2421022 & 97 & .28 & -- \\
X13066   & 6-8-1916     & 2421023  & 90  & .28 & --\\
X13070   & 6-19-1916    & 2421034  & 101 & .28 & $\eta$ Car not on plate \\
X13072\tablenotemark{b}   & 6-20-1916    & 2421035  & 83  & .29 & --  \\ 
X13358   & 5-5-1922     & 2423180  & 60  & .35 &  -- \\
X13365   & 5-10-1922    & 2423185  & 61  & .36 &  -- \\
X13370   & 5-15-1922    & 2423190  & 56  & .36  & unwidened \\
X13371\tablenotemark{b}   & 5-15-1922    & 2423190  & 60  & .36  & --  \\ 
X13373   & 5-16-1922    & 2423191  & 60  & .36 &  -- \\
X13376   & 5-17-1922    & 2423192  & 64  & .36 & -- \\
X13396   & 6-1-1922     & 2423209  & 60  & .36 & -- \\
X13408\tablenotemark{b}   & 6-12-1922    & 2423218  & 60  & .37 & -- \\
X13410   & 6-13-1922   & 2423219   & 60  &  .37  & -- \\
X13412   & 6-14-1922   & 2423220   & 70  &  .37  & missing \\
X13413\tablenotemark{b} & 6-15-1922 & 2423221 & 64 & .37 & -- \\
X13692\tablenotemark{b} & 4-16-1923   &  2423526 & 57 & .52 & -- \\
X13739   & 5-10-1923    & 2423550  &  85 &  .54 & missing \\
X14861   & 5-3-1926     & 2424639  & 50  & .07  &  -- \\
X14939   & 6-15-1926    & 2424682  & 20  & .09  & -- \\
X14961\tablenotemark{b} & 7-1-1926 & 2424698 & 40 &  .10 & -- \\
X16486  & 3-11-1937    & 2428604  & 120 &  .03 &  missing? \\
X16492\tablenotemark{b}  & 5-2-1937  &  2428656 &  45 & .06  & -- \\
X16493  & 5-2-1937     & 2428656 & 20 &  .06  & poor spectrum \\
X16499\tablenotemark{b}  &  5-28-1937 & 2428682 & 45 & .07 & -- \\
X16502\tablenotemark{b} & 5-31-1937 & 2428685  & 120 &  .07  & -- \\
X16503\tablenotemark{b} & 5-31-1937 & 2428685  & 49 & .07 &  -- \\
X16506  & 6-6-1937  & 2428691 & 120 &  .07 & -- \\
X16507  & 6-6-1937  & 2428691 & 45  & .07  & -- \\
X16511  & 6-7-1937  & 2428692 & 120 & .07  & -- \\
X16512  & 6-7-1937  & 2428692 & 45  & .07 & -- \\
X16516  & 6-10-1937 & 2428692 & 120 & .08  & -- \\
X16517  & 6-10-1937 & 2428692 & 45  &  .08  & -- \\
X16521  & 6-11-1937 & 2428696 & 20 & .08  & -- \\
X16531  & 6-29-1937 & 2428714 & 20 & .09 & poor spectrum \\
X16602  & 3-20-1938 & 2428978 & 45 & .22 & rejected plate \\
X16605  & 3-21-1938 & 2428979 & 120 & .22 & poor spectrum \\
X16610  & 3-27-1938 & 2428985 &  70 & .22  & missing \\
X16614\tablenotemark{b} & 3-29-1938 & 2428987 & 55 & .22 & good spectrum \\
X16618\tablenotemark{b} & 4-10-1938 & 2428999 & 40 &  .23  &  -- \\
X16664  & 1-2-1939 & 2429266 & 45  & .36  & missing \\
X16676  & 2-24-1939 & 2429319 & 120 & .38 & rejected plate \\
X16677  & 2-24-1939 & 2429319 & 20  & .38 & --\\
X16684  & 2-27-1939 & 2429322 & 4  & .38  & rejected plate \\
X16685  & 2-27-1939 & 2429322 & 45  & .38  &  -- \\
X16686\tablenotemark{b} & 2-27-1939 & 2429322 & 20 & .39 & -- \\
X16694  & 3-12-1939 & 2429335 & 45 & .39  & fuzzy image \\
X16695  & 3-12-1939 & 2429335 & 20 &  .39 & -- \\
X16698  & 3-16-1939 & 2429339 & 20 &  .39 & -- \\
X16701  & 3-17-1939 & 2429340 & 120 & .39 & -- \\
X16704  & 3-19-1939 & 2429342 & 45 &  .39 & -- \\
X16707  & 3-20-1939 & 2429343 & 47 & .39  & -- \\
X16708  & 3-24-1939 & 2429347 & 105 & .39 & $\eta$ Car not on plate \\
X16710  & 3-26-1939 & 2429349 & 120 & .39  & -- \\
X16715\tablenotemark{b}  & 4-10-1939 & 2429364 & 45 & .40 & -- \\
X16720  & 4-16-1939 & 2429370 & 120 & .40 & -- \\
X16724  & 4-19-1939 & 2429373 & 120 & .41 & -- \\
X16729  & 4-20-1939 & 2429374 & 45 & .41  & -- \\
X17165\tablenotemark{b} & 5-1-1940 & 2429751 & 120 & .60 & -- \\
X17170  & 5-6-1940 & 2429756 & 120 & .60 & -- \\
X17174  & 5-7-1940 & 2429757 & 120 & .60 & fuzzy image \\
X17180  & 5-10-1940 & 2429760 & 120 & .60 & -- \\
X17421  & 2-23-1941 & 2430049 & 120 & .75 & -- \\
X17433  & 4-15-1941 & 2430100 & 45 & .75  & -- \\
X17437\tablenotemark{b} & 4-21-1941 & 2430106 & 120 & .77 & -- \\
X17440\tablenotemark{b} & 4-24-1941 & 2430109 & 45 & .77 & --\\  
X17444  & 5-1-1941  & 2430116 & 120 & .78  & -- \\
X17465\tablenotemark{b} & 5-21-1941 & 2430136 & 120 & .79  & -- \\
X17466 & 5-21-1941 & 2430136 & 35 & .79 & -- \\
X17472 & 5-23-1941 & 2430138 &  120 & .79 & -- \\
X17473 & 5-23-1941 & 2430138 & 45 & .79 & -- \\
X17476 & 5-25-1941 & 2430140 & 120 & .79 & -- \\
X17477 & 5-25-1941 & 2430140 &  45 & .79 & -- \\
X17481 & 5-26-1941 & 2430141 & 120 & .79 & -- \\
X17482\tablenotemark{b} & 5-26-1941 & 2430141 & 45 & .79 & -- \\ 
\enddata
\tablenotetext{a}{weak absorption line spectrum; H$\beta$, H$\gamma$, H$\delta$, [Fe II] 4359{\AA} emission lines.}
\tablenotetext{b}{digitized} 
\tablenotetext{c}{absorption lines barely discernible; H$\beta$, H$\gamma$ emission}
\tablenotetext{d}{absorption lines very weak or gone; H$\beta$, H$\gamma$, H$\delta$, [Fe II
4359{\AA} emission.}
\tablenotetext{e}{Hydrogen, Fe II, [Fe II] emission} 
\end{deluxetable}


\end{document}